\documentclass[aip,jcp,graphicx,reprint]{revtex4-1} 
\usepackage{physics}
\usepackage{graphicx} 
\graphicspath{{Materials/}} 
\usepackage{amsmath,amssymb,amsthm} 
\usepackage{dcolumn} 
\usepackage{tikz}
\usepackage[font={small,it}]{caption}
\usepackage[font={small,rm}]{subcaption}
\usepackage[hidelinks]{hyperref}
\DeclareMathSizes{10}{8}{7}{6} 
\usepackage{appendix}
\usepackage{cleveref}
\usepackage{tikz}
\usetikzlibrary{decorations.markings, shapes.geometric, positioning,  intersections, 3d,calc,math}
\usepackage{algorithm}
\usepackage{algorithmic}
\usepackage{bm} 
\usepackage{placeins}
\usepackage{hhline}
\usepackage{multirow}

\begin{document}
\title{
Transition Path and Interface Sampling of Stochastic Schr\"{o}dinger Dynamics
}
\author{Robson Christie}
\affiliation{Department of Mathematics, Imperial College London,
United Kingdom}

\author{Peter G. Bolhuis }
\affiliation{van 't Hoff Institute for Molecular Sciences, University of Amsterdam, Netherlands}
\author{David T. Limmer}
\affiliation{%
 Department of Chemistry, University of California, Berkeley, CA 94720, USA
}%
\affiliation{%
 Kavli Energy NanoScience Institute, Berkeley, CA 94720, USA
}%
\affiliation{%
 MSD, Lawrence Berkeley National Laboratory, Berkeley, CA 94720, USA
}%
\affiliation{%
 CSD, Lawrence Berkeley, National Laboratory, Berkeley, CA 94720, USA
}%

\begin{abstract}
We study rare transitions in Markovian open quantum systems driven with Gaussian noise, applying transition path and interface sampling methods to trajectories generated by stochastic Schr\"{o}dinger dynamics. Interface and path sampling offer insights into rare event transition mechanisms 
while simultaneously establishing a quantitative measure of the associated rate constant.  Here, we extend their domain to systems described by stochastic Schrödinger equations. As a specific example, we explore a model of quantum Brownian motion in a quartic double well, consisting of a particle coupled to a Caldeira-Leggett oscillator bath, where we note significant departures from the Arrhenius law at low temperatures due to the presence of an anti-Zeno effect. 

\end{abstract}

\maketitle 

\section{Introduction}
Metastability in open quantum systems represents a frontier area of quantum mechanics due to its relevance to information processing and computing. 
Rare events that relax metastable systems can destroy coherence, resulting in the loss of information or can induce transitions between different chemical states in molecules\cite{ref:chandlerbook}. In classical complex molecular systems, rare events are connected to high free energy barriers and improbable configurational fluctuations, but in quantum systems tunneling and interference can also play a large role.\cite{topaler1994quantum,pollak2024instanton} Since straightforward simulations are often unfeasible due to the long timescales involved, a variety of importance sampling methods have been developed to study such infrequent events.\cite{tokdar2010importance} Transition path sampling (TPS)\cite{bolhuis2002transition} and its extensions, notably transition interface sampling(TIS) \cite{van2005elaborating}  have in particular played an important role in elucidating rare events in complex, classical systems\cite{dellago1998efficient,basner2005enzyme,bolhuis2003transition,varilly2013water} and have been extended to quantum jump processes.\cite{schile2018studying} 
TPS algorithms collect an ensemble of reactive pathways that connect a predefined initial and final state (e.g. reactant and product), which can be scrutinised for mechanistic detail and evaluation of observables such as time correlation functions that yield the reaction rate constant\cite{bolhuis2002transition}. Path sampling has the advantage of being agnostic to the nature of the transition state region.
However,  application of path sampling methods to open quantum systems has been limited, as many of the formalisms used to simulate open quantum systems do not rigorously preserve the required detailed balance. Here, we extend path sampling tools to open quantum systems described by a stochastic Schrödinger equation \cite{percival,belavkin1989nondemolition},  by deriving its associated path action and constructing efficient trajectory moves for a system without microscopic reversibility. 



This extension enables the exploration of quantum transition processes that are otherwise computationally intractable due to their rarity. We explore a model of quantum Brownian motion comprising a quartic double well with a particle coupled to a thermal Caldeira-Leggett oscillator bath \cite{caldeira1983path}, demonstrating deviations from the classical Arrhenius law at low temperatures due to the presence of an anti-Zeno effect \cite{AntiZeno1,AntiZeno2,maniscalco2006zeno}. This choice of model serves as a proof of concept for rare-event sampling of the stochastic Schrödinger equation with the possibility for future applications in more complex quantum molecular systems. Our approach complements studies applying path sampling to open quantum systems driven by discrete Poissonian noise \cite{schile2018studying,anderson2022nonadiabatic} and other techniques utilized to sample quantum transition paths  \cite{reiner2023nonadiabatic}. To highlight the potential applications of the path and interface sampling of quantum Brownian trajectories, we will present our findings alongside results obtained from the widely used \cite{dellago1998transition,moroni2004investigating,delarue2017ab,orland2011generating,jung2017transition} classical implementation of path and interface sampling for the Langevin equation.

This paper is organised as follows: In the next section we introduce the theoretical background for classical and quantum stochastic dynamics and derive the path sampling algorithms for these equations of motion.
In section 3 we illustrate our novel algorithms on a simple quartic double well potential. We end with concluding remarks.

\section{Theoretical background}
\subsection{Classical and quantum Langevin equations}

The Langevin equation is extensively used in molecular dynamics simulations\cite{paquet2015molecular}, as a model of a dissipative environment, incorporating the influence of random forces and friction. For a particle with mass $m$ in one dimension, the resulting dynamics take the following form \cite{pavliotis2014stochastic}
\begin{equation} \label{eq:Langevin_tilde}
\begin{pmatrix}
dx\\ dp
\end{pmatrix}=\left [\frac{1}{m}\begin{pmatrix} p\\0 \end{pmatrix}-\begin{pmatrix}0 \\ V'(x)\end{pmatrix}-2\gamma \begin{pmatrix} 0 \\ p \end{pmatrix} \right ] dt+\begin{pmatrix}
0\\2 \sqrt{ \gamma m k_B T }
\end{pmatrix}d\xi,
\end{equation}
where \(d\xi\) represents a real It\^{o} process \cite{bernt}, and \(x\) and \(p\) denote the position and momentum of the particle, respectively.
Furthermore, $V(x)$ denotes the potential energy as a function of $x$,  $t$ is the time,   $\gamma$ is the friction coefficient, $k_B$ is Boltzmann's constant, and $T$ is the temperature of the heat bath. 
Dual to the Langevin equation is the Fokker-Planck master equation of the phase space probability density, $\rho(x,p)$ of the Langevin particle
\begin{multline} \label{eq:Fokker}
	\frac{\partial}{\partial t}\rho(x,p)=-\frac{p}{m} \frac{\partial}{\partial x}\rho(x,p)+V'(x)\frac{\partial}{\partial p}\rho(x,p)\\ +  2 \gamma \frac{\partial}{\partial p}\left(p \rho(x,p) \right) + 2\sqrt{m \gamma k_B T}\frac{\partial^2}{\partial p^2} \rho(x,p).
\end{multline}
A significant contribution to the theory of quantum Brownian motion may be found in the seminal work of Caldeira and Leggett \cite{caldeira1983path}, in which they derived the following master equation for the reduced density matrix $\hat \rho$ of a quantum particle coupled to a thermal oscillator bath
\begin{equation} \label{eq:Caldeira-Master}
\frac{d}{dt} \hat \rho = -\frac{i}{\hbar}[\hat H, \hat \rho]-\frac{i \gamma}{\hbar}[\hat X \{\hat P, \hat \rho\}]-\frac{2 m \gamma k_B T}{\hbar}[\hat X, [\hat X, \hat \rho]],
\end{equation}
with $[.,.]$ the commutator and $\hat H$ the Hamiltonian operator
\begin{equation}
    \hat H=\frac{\hat P^{2}}{2 m}+V(\hat X),
\end{equation}
with $\hat X,\hat P$ the position and moment operators respectively. Furthermore, $h$ denotes Planck's constant, and $\{.,.\}$ represents the anti-commutator.

It is well known that the evolution in Eq. \eqref{eq:Caldeira-Master} does not preserve the positivity of the density matrix at low temperatures\cite{diosi1993calderia,homa2019positivity}. To address this one can make a modification to \eqref{eq:Caldeira-Master} which consists of adding the term \cite{petruccione}
\begin{equation}
-\frac{\gamma}{8 m k_B T}[\hat P,[\hat P,\hat \rho]].
\end{equation}
proportional to the friction. This modification term is negligible in the high-temperature limit, but allows the reformulation of Eq.~\eqref{eq:Caldeira-Master} in the positivity-preserving {Lindblad form} \cite{lindblad1976generators,gorini1976completely}
\begin{equation} \label{eq:lindblad}
\frac{d}{dt}\hat{\rho}=-\frac{i}{\hbar}[\hat{H}_{\gamma},\hat{\rho}] + \frac{1}{\hbar} \left(\hat{L} \hat{\rho} \hat{L}^{\dagger} -\frac{1}{2} \hat{L}^{\dagger} \hat{L}\hat{\rho}-\frac{1}{2}\hat{\rho} \hat{L}^{\dagger} \hat{L}\right),
\end{equation}
with
\begin{equation} \label{eq:heatbathH}
\hat{H}_{\gamma}=\frac{\hat{P}^2}{2m}+V(\hat{X})+\frac{\gamma}{2}(\hat{X}\hat{P}+\hat{P}\hat{X})
\end{equation}
and
\begin{equation} \label{eq:heatbathL}
    \hat{L}= \sqrt{\frac{4 \gamma m k_B T}{\hbar}}\hat{X}+i\sqrt{\frac{\gamma \hbar}{4mk_B T}}\hat{P},
\end{equation}
and where  $\hat L^\dagger$ denotes the complex conjugate of $\hat L$. 
The above expressions are often regarded as a quantum counterpart to the classical Fokker-Planck equation, Eq.~\eqref{eq:Fokker}. Analogously, a quantum counterpart to the Langevin equation \eqref{eq:Langevin_tilde} may be given in the form of a stochastic Schr\"{o}dinger equation (SSE)\cite{percival,belavkin1989nondemolition}, which describes the time evolution of the wave function $\psi$ as 
\begin{multline}
\label{eq:SSE}	    
    \ket{d \psi } = \frac{1}{\hbar}\left(-{i}\hat{H}_{\gamma} -\frac{1}{2 }\hat{L}^{\dagger}\hat{L}+\ev*{\hat{L}^{\dagger}}\hat{L} - \frac{1}{2}\ev*{\hat{L}^{\dagger}}\ev*{\hat{L}}\right)\ket {\psi} dt\\ +\frac{1}{\sqrt{ \hbar}}\left(\hat{L}-\ev*{\hat{L}}\right)\ket{\psi} d\xi.
\end{multline}
where the angular brackets denote an expectation value. We will use this model extensively in \cref{sec:results}.

\subsection{Path and Interface Sampling}
Provided the stochastic Schr\"{o}dinger equation, we can define an ensemble of stochastic trajectories and in principle apply importance sampling methods like   \textit{Transition Path Sampling} (TPS)\cite{bolhuis2002transition} and \textit{Transition Interface Sampling}  (TIS)\cite{van2005elaborating}.
TPS generates a set of reactive pathways, those conditioned to start and end in particular states, by perturbing the trajectory at a selected ``\textit{shooting point}'' along a given path and integrating equations of motion forward and backward in time. These paths can then be used to calculate correlation functions and through reversible work analogues, the rate. TIS uses a shooting algorithm with a variable path length and calculates the rate by dividing the transition into multiple stages, improving the likelihood of sampling acceptance at the cost of introducing a progress coordinate.

Because the SSE does not strictly obey detailed balance, its dynamics are not microscopically reversible. In cases where the dynamics are not time-reversible, the acceptance rule for a shooting move must be adapted from its original formulation to exclude backward segments with minimal probability of occurrence. To address this, TPS and TIS employ the Metropolis acceptance probability to validate newly generated paths, ensuring convergence of the Markov chain of paths to the correct ensemble in path space \cite{metropolis1953equation,dellago1998transition}. This acceptance probability for any path move is generically given by
\begin{equation}
\begin{aligned}
\mathcal{P}_{\text {acc }}\left[\Psi^{(\mathrm{o})} \rightarrow\right. & \left.\Psi^{(\mathrm{n})}\right]=h_A\left[x^{(\mathrm{n})}(t_0)\right] h_B\left[x^{(\mathrm{n})}(t_f)\right] \\
& \times \min \left\{1, \frac{\mathcal{P}\left[\Psi^{(\mathrm{n})}\right] \mathcal{P}_{\mathrm{gen}}\left[\Psi^{(\mathrm{n})} \rightarrow \Psi^{(\mathrm{o})}\right]}{\mathcal{P}\left[\Psi^{(\mathrm{o})}\right] \mathcal{P}_{\mathrm{gen}}\left[\Psi^{(\mathrm{o})} \rightarrow \Psi^{(\mathrm{n})}\right]}\right\},
\end{aligned}
\end{equation}
where $\Psi^{(\mathrm{o})}$, $\Psi^{(\mathrm{n})}$, denote the old and new trajectory, respectively, $h_A[x]$ and $h_B[x]$  are indicator functions that are unity if $x\in A $ or $x\in B $, respectively and zero otherwise. These functions act on the trajectories' initial and final configurations. The path probability is denoted
$\mathcal{P}[\Psi]$. In appendix \ref{sec:SSE-Derivation} we derive explicit expressions for the path probability based on the  SSE. 
In the case of a displacement at the shooting point $t_s$ followed by a forward and backward shot, the acceptance criteria reduces to 
\begin{multline} \label{eq:PaccShoot}
     \mathcal{P}_{\mathrm{acc}}\left[\Psi^{(\mathrm{o})} \rightarrow \Psi^{(\mathrm{n})}\right]=\\h_A\left[x_{t_0}^{(\mathrm{n})}\right] h_B\left[x_{t_f}^{(\mathrm{n})}\right] \times \min \Bigg[1, \frac{\rho_{st}\left(\psi_{t_0}^{(\mathrm{n})}\right)\mathcal{P}_{\mathrm{gen}}\left[\psi_{t_s}^{(\mathrm{n})} \rightarrow \psi_{t_s}^{(\mathrm{o})}\right]}{\rho_{st}\left(\psi_{t_0}^{(\mathrm{o})}\right)\mathcal{P}_{\mathrm{gen}}\left[\psi_{t_s}^{(\mathrm{o})} \rightarrow \psi_{t_s}^{(\mathrm{n})}\right]} \\ \prod_{i=0}^{s} \frac{\mathcal{P}\left(\psi_{t_i}^{(\mathrm{n})} \rightarrow \psi_{t_{i+1}}^{(\mathrm{n})}\right)}{\bar{\mathcal{P}}\left(\psi_{t_{i+1}}^{(\mathrm{n})} \rightarrow \psi_{t_i}^{(\mathrm{n})}\right)} \times \frac{\bar{\mathcal{P}}\left(\psi_{t_{i+1}}^{(\mathrm{o})} \rightarrow \psi_{t_i}^{(\mathrm{o})}\right)}{\mathcal{P}\left(\psi_{t_i}^{(\mathrm{o})} \rightarrow \psi_{t_{i+1}}^{(\mathrm{o})}\right)}\Bigg],
\end{multline}
with $\mathcal{P}$ and $\bar{\mathcal{P}}$ denoting the forward and backward transition probabilities respectively, $\rho_{st}$ is the stationary distribution of the corresponding master equation (Fokker Planck or Lindblad). Typically, the displacement at the shooting point is selected from a symmetric distribution, such that
\begin{equation}\mathcal{P}_{\mathrm{gen}}\left[\psi_{t_s}^{(\mathrm{o})} \rightarrow \psi_{t_s}^{(\mathrm{n})}\right] = \mathcal{P}_{\mathrm{gen}}\left[\psi_{t_s}^{(\mathrm{n})} \rightarrow \psi_{t_s}^{(\mathrm{o})}\right],
\end{equation}
ensuring that the ratio in \eqref{eq:PaccShoot} becomes unity.
\begin{figure}
\begin{tikzpicture}[scale=0.7, decoration={markings, mark=at position 0.2 with {\arrow{>}}, mark=at position 0.4 with {\arrow{>}}, mark=at position 0.7 with {\arrow{>}}, mark=at position 0.9 with {\arrow{>}}}]
\draw[thick,->] (-2,0,0) -- (8,0,0) node[right]{$t$};
\draw[thick,->] (0,-2,0) -- (0,2,0) node[above]{$p$};
\draw[thick,->] (0,0,-2) -- (0,0,2) node[below left]{$q$};
\foreach \i/\j in {0/$t_0$ , 1.5/, 3/$t_s$ , 4.5/, 6/$t_N$} {
    \fill[red, opacity=0.3] (\i,2,-2) -- (\i,2,-1) -- (\i,-2,-1) -- (\i,-2,-2) -- cycle;
    \fill[gray, opacity=0.1] (\i,2,-1) -- (\i,2,1) -- (\i,-2,1) -- (\i,-2,-1) -- cycle;
    \fill[blue, opacity=0.3] (\i,2,1) -- (\i,2,2) -- (\i,-2,2) -- (\i,-2,1) -- cycle;
    \ifx\j\empty\else
        \node at (\i-0.5, -3, 0) {\j};
    \fi
}
\draw[postaction={decorate}, red, thick, domain=0:6, samples=10, smooth] plot (\x, {cos(deg(0.5*\x))^2}, {-1.3*cos(deg(0.5*\x))}) node[ right]{$\Psi^{o}(t)$};
\foreach \i in {1.5,3,4.5} {
    \fill[red] (\i, {cos(deg(0.5*\i))^2}, {-1.3*cos(deg(0.5*\i))}) circle (1.5pt);
}
\draw [fill=white] (0, {cos(deg(0.5*0))^2}, {-1.3*cos(deg(0.5*0))}) circle [radius=.07];
\fill[black] (6, {cos(deg(0.5*6))^2}, {-1.3*cos(deg(0.5*6))}) circle (1.5pt);

\draw[postaction={decorate}, blue, thick, domain=0:6, samples=10, smooth] plot (\x, {1.2*sin(deg(0.5*\x))^2}, {-1.7*cos(deg(0.5*\x))}) node[ right]{$\Psi^{n}(t)$};
\foreach \i in {1.5,3,4.5} {
    \fill[blue] (\i, {1.2*sin(0.5*deg(\i))^2}, {-1.7*cos(deg(0.5*\i))}) circle (1.5pt);
}
\draw [fill=white] (0, {1.2*sin(0.5*deg(0))^2}, {-1.7*cos(deg(0.5*0))}) circle [radius=.07];
\fill[black] (6, {1.2*sin(0.5*deg(6))^2}, {-1.7*cos(deg(0.5*6))}) circle (1.5pt);

\draw[black, ->, thick] 
  (3, {cos(deg(0.5*3))^2+0.1}, {-1.3*cos(deg(0.5*3))}) -- 
  (3, {1.2*sin(deg(0.5*3))^2-0.1}, {-1.7*cos(deg(0.5*3))})
    node[pos=0.1, right, font=\scriptsize] {$\Delta p$};
\end{tikzpicture}
    \caption{Two way shooting with path adjustment: The new trajectory $\Psi^{n}$ (blue) is generated by modifying the momentum of $\Psi^{o}$ (red) at $t=t_s$ the dynamics are then integrated forward and backward.}
    \label{fig:2-Way-Shooting}
\end{figure}
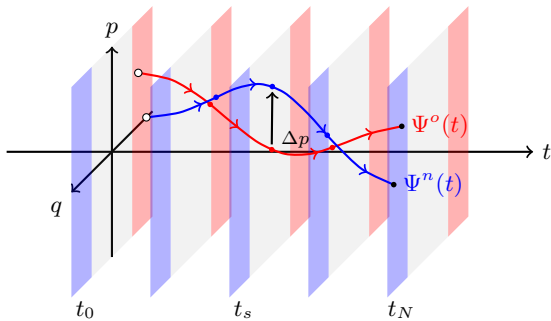

Often shooting moves in TPS are sufficient for calculating correlation functions and rates, but in non-time-reversal symmetric systems these can have low acceptance probabilities. To remedy this, we present \textit{Mirror moves}, a novel variant of this method. Mirror TPS moves utilize involutory (self-inverse) transformations to efficiently generate multiple proposal paths simultaneously, minimizing the use of computational resources. Additionally, this method enhances the decorrelation process between successive trajectories. The detailed methodology of Mirror TPS is elaborated in \cref{app:MirrorTPS}.

The correlation function $C(t)$ plays a pivotal role in understanding the timescales of the transition dynamics and is given as
\begin{equation} \label{eq:Correl}
C(t) \equiv \frac{\left\langle h_A\left(x_0\right) h_B\left(x_t\right)\right\rangle}{\left\langle h_A\right\rangle}.
\end{equation}
Following \cite{bolhuis2002transition} this can be expressed in terms of the ratio
\begin{equation}
    C(t)=\frac{\left\langle h_B\left(x_t\right)\right\rangle_{A B}^*}{\left\langle h_B\left(x_{t^{\prime}}\right)\right\rangle_{A B}^*} \times C\left(t^{\prime}\right)
\end{equation}
where the notation $\left\langle h_B\left(x_t\right)\right\rangle_{A B}^*$ indicates an average over the ``\textit{visiting}" ensemble of pathways that start in $A$ but are only required to visit $B$ at some time slice along the path rather than end in $B$. We thus calculate the full correlation function at $C(t')$ using umbrella sampling\cite{bolhuis2002transition} to calculate a scale factor followed by a visiting ensemble simulation for the rest of the time dependence of $C(t)$.

TPS can be used to calculate the rate constant by determining the time derivative of the correlation function in the region where it increases linearly. A more efficient calculation of the rate constant can be achieved using TIS. TIS operates by dividing the transition from a reactant state, denoted as \( A \), to a product state, \( B \), into a series of interfaces. These interfaces are positioned between the initial and final states and are successively crossed by a trajectory moving from \( A \) to \( B \). The rate calculation in TIS is then broken down into two main components:
\begin{itemize}
    \item Flux calculation across the first interface: This involves counting the number of times trajectories originating from the reactant state \( A \) cross the first interface in the outward direction per unit time. This flux $\Phi_0$ quantifies how often the system attempts to make a transition.
    \item Interface sampling: This computes the overall probability of crossing by multiplying the conditional probabilities \( \mathcal{P}(\lambda_{i+1} | \lambda_i) \) which describe the likelihood of a trajectory crossing the subsequent interface, contingent on it having already crossed the current one.
\end{itemize}
The product of the total crossing probability and the attempt frequency then gives the rate constant as
\begin{equation}\label{eq:TISRate}
    k_t=\Phi_0\prod_{i}\mathcal{P}(\lambda_{i+1}|\lambda_i).
\end{equation}
In the next section, we recap how to implement path sampling for Langevin dynamics as this will be later contrasted with a quantum implementation in \cref{sec:QTPS}.

\subsection{Classical Langevin dynamics} \label{sec:CTPS}

In one dimension, a classical trajectory with $N$ timesteps can be characterized as a $2\times N$ matrix
\begin{equation} \Psi =
\begin{pmatrix}
    \psi_{1} & \psi_{2} & \cdots & \psi_{N}
\end{pmatrix}=
\begin{pmatrix}
    x_{1} & x_{2} & \cdots & x_{N} \\
    p_1 & p_2 & \cdots & p_N \\
\end{pmatrix},
\end{equation}
with the system state at time $t_j$ given by $\psi_j$. In the 1-d case, we can define our reaction coordinate as $x$ and thus the reaction path is given by the top row of $\Psi$
\begin{equation}
    \Gamma=\Psi(1,:)=\begin{pmatrix}
    x_{1} & x_{2} & \cdots & x_{N}
\end{pmatrix}.
\end{equation}
To generate a new path we perturb the momentum at the shooting point $\psi_S$ at $t_S$:$\psi_S\to D_{\delta p}(\psi_S)$ , where $D_{\delta p}$ is the operator that performs the perturbation. We then integrate the Langevin dynamics forward for $j=S+1:N$ with initial state $D_{\delta p}(\psi_S)$ generating $\Psi_F$. We can perform the backward evolution by evolving the time-reversed state $\mathbb{T} (D_{\delta p}(\psi_S)))$ forward in time from $j=1:S-1$ generating $\Psi_B$ then time reversing the resulting segment. The newly generated trajectory is given by
\begin{equation}
\Phi=
\begin{pmatrix} \label{eq:time_rev_trick}
    \mathbb{T}(\Psi_B)& D_{\delta p}\psi_S &\Psi_F 
\end{pmatrix}.
\end{equation}
In terms of position and momentum, the time reversal operator acting on the classical trajectory $\Psi$ flips the order and changes the sign of the momentum as follows
\begin{equation}
    \mathbb{T}\Psi=\mathbb{T}\begin{pmatrix}
    x_{1} & x_{2} & \cdots & x_{N} \\
    p_1 & p_2 & \cdots & p_N \\
\end{pmatrix}=\begin{pmatrix}
    x_{N} & x_{N-1} & \cdots & x_{1} \\
    -p_N & -p_{N-1} & \cdots & -p_1 \\
\end{pmatrix}.
\end{equation}
For later use with Mirror TPS, we also define the parity operator $\mathbb{P}$ acting on this trajectory as 
\begin{equation}
    \mathcal{P} \Psi =\begin{pmatrix}
   - x_{1} & -x_{2} & \cdots & -x_{N} \\
    -p_1 & -p_2 & \cdots & -p_N \end{pmatrix}=-\Psi.
\end{equation}
To calculate the acceptance probabilities \eqref{eq:PaccShoot} of generated classical paths we can substitute the Langevin drift and diffusion coefficients
\begin{equation}
    u(\Psi)=\begin{pmatrix}
        \frac{p}{m}\\ -V'(x)-2 \gamma p
    \end{pmatrix}\quad \text{and}\quad\sigma=\begin{pmatrix}
        0\\ 2\sqrt{\gamma m k_B T}.
    \end{pmatrix}
\end{equation},
into the expression for the single-step transition probability \eqref{eq:P_Step} derived in \cref{sec:SSE-Derivation}, this yields
\begin{multline}
    \mathcal{P}\left( \psi'_{0} \to \psi''_{\delta t}\right)=\\\frac{1}{\sqrt{2 \pi \delta t}}\exp{-\frac{ \left(p''-p'+\delta t (V'(x')+2\gamma p') \right)^2}{8 m\gamma k_B T \delta t}}.
\end{multline}
The above proportionality constant cancels when calculating acceptance probabilities using Metropolis rule \eqref{eq:PaccShoot}.

\subsection{Stochastic Quantum Dynamics} \label{sec:QTPS}

Having recapped the classical case, we now describe how to implement path sampling for the SSE \eqref{eq:SSE} with Lindblad operator \eqref{eq:heatbathL}. We find that the state re-localizes as an approximately Gaussian state \cite{halliwell1995quantum} in one of the metastable wells shortly after each transition. This allows us to use the position expectation value $\langle \hat{X} \rangle$ as a reaction coordinate.

In the quantum case, we must choose a basis in which to express our states and observables; a convenient choice is the Fock basis of harmonic oscillator eigenstates $\{\ket{n}\}$. To generate a candidate trajectory via shooting, we proceed with the same steps as in the classical case by selecting a shooting point, perturbing the momentum, and integrating forward and backward in time. To perturb the momentum at the shooting point, we apply a random momentum displacement operator, defined as:
\begin{equation}
\hat{D}(\delta p) = e^{\delta p \hat{X}},
\end{equation}
where $\delta p$ is selected from a symmetric distribution. The time reversal operator acting on a quantum trajectory is simply order reversal and complex conjugation:
\begin{equation}
    \mathbb{T} \begin{pmatrix} \psi_{1} & \psi_{2} & \cdots & \psi_{N} \end{pmatrix} = \begin{pmatrix} \psi_{N}^{*} & \psi_{N-1}^{*} & \cdots & \psi_{1}^{*} \end{pmatrix}.
\end{equation}
To propagate the SSE trajectory forward in time, we employ the stochastic Euler algorithm \ref{alg:Euler-NSSE}, which incorporates a renormalization step at each timestep. Although the SSE conserves the norm of the state in the continuum limit, discretization introduces small numerical errors that can lead to a gradual loss of norm. This is because the selection rule \eqref{eq:PaccShoot} is inherently biased towards trajectories with a norm less than unity. To calculate the acceptance probability, we use \eqref{eq:P_StepC} derived in \cref{sec:SSE-Derivation} to express the single-step transition probability for the SSE as:
\begin{multline}\label{eq:P_StepText}
    \mathcal{P}\left( \psi'_{0} \to \psi''_{\delta t}\right) \\= \frac{1}{\sqrt{2 \pi \delta t}} \exp{-\frac{\left|\sigma^{\dagger}(\psi')\left(\psi''-\psi'- u(\psi')\delta t\right)\right|^2}{2 \delta t (\sigma^{\dagger}(\psi') \sigma(\psi'))^2}}
\end{multline}
with drift and diffusion vectors given by:
\begin{align}
    u(\Psi) &= \frac{1}{\hbar}\left(-i \hat{H}_{\gamma} -\frac{1}{2} \hat{L}^{\dagger}\hat{L}+\langle \hat{L}^{\dagger} \rangle \hat{L} - \frac{1}{2} \langle \hat{L}^{\dagger} \rangle \langle \hat{L} \rangle \right) \ket {\psi} \\
    \sigma(\Psi) &= \frac{1}{\sqrt{\hbar}}\left(\hat{L}-\langle \hat{L} \rangle\right) \ket{\psi}.
\end{align}
The stationary distribution in the acceptance rule \eqref{eq:PaccShoot} can be found by writing the Lindblad superoperator in matrix form and numerically finding the null vector \cite{explind,campaioli2024quantum}.

\begin{algorithm}[H]
\caption{SSE stochastic Euler propagation with stepwise renormalization}
\begin{minipage}{0.95\linewidth} 
\begin{algorithmic}[1] \label{alg:Euler-NSSE}
	\STATE Discretise the time interval $(t_f,t_0) \rightarrow \{t_k \} $ with $\Delta t=t_{k+1}-t_k$.
	\STATE  Define two functions corresponding to the deterministic and stochastic components of the SSE \eqref{eq:SSE},
	\begin{align*}
		u(\ket{\psi(t_k)})&= \frac{1}{\hbar}\left(-i\hat{H}_{\gamma} -\frac{1}{2}\hat{L}^{\dagger}\hat{L}+\ev*{\hat{L}^{\dagger}}\hat{L} - \frac{1}{2}\ev*{\hat{L}^{\dagger}}\ev*{\hat{L}}\right)\ket{\psi(t_k)} \\
		\sigma(\ket{\psi(t_k)})&=  \frac{1}{\sqrt{ \hbar}}(\hat{L}-\ev*{\hat{L}}) \ket {\psi(t_k)}.
	\end{align*}
where $\xi$ is a $\mathcal{N}(0,1)$ distributed real random variable.
	\STATE Calculate the next timestep as 
\begin{equation}
		\ket{\psi(t_{k+1})}\!=\!\frac{1}{\norm{\psi(t_k)}}\left(\ket{\psi(t_k)}\!+\!u(\ket{\psi(t_k)})\Delta t+\sigma(\ket{\psi(t_k)})\xi\sqrt{\Delta t}\right).
\end{equation}
	\STATE Repeat steps $3\;\&\;4$ or all $k$.
\end{algorithmic}
\end{minipage}
\end{algorithm}
\section{Results and Discussion} \label{sec:results}
\begin{figure*}
\centering 
\begin{tabular}{c c c}
\begin{subfigure}[c]{.3\textwidth} 
	  \includegraphics[width=\textwidth]{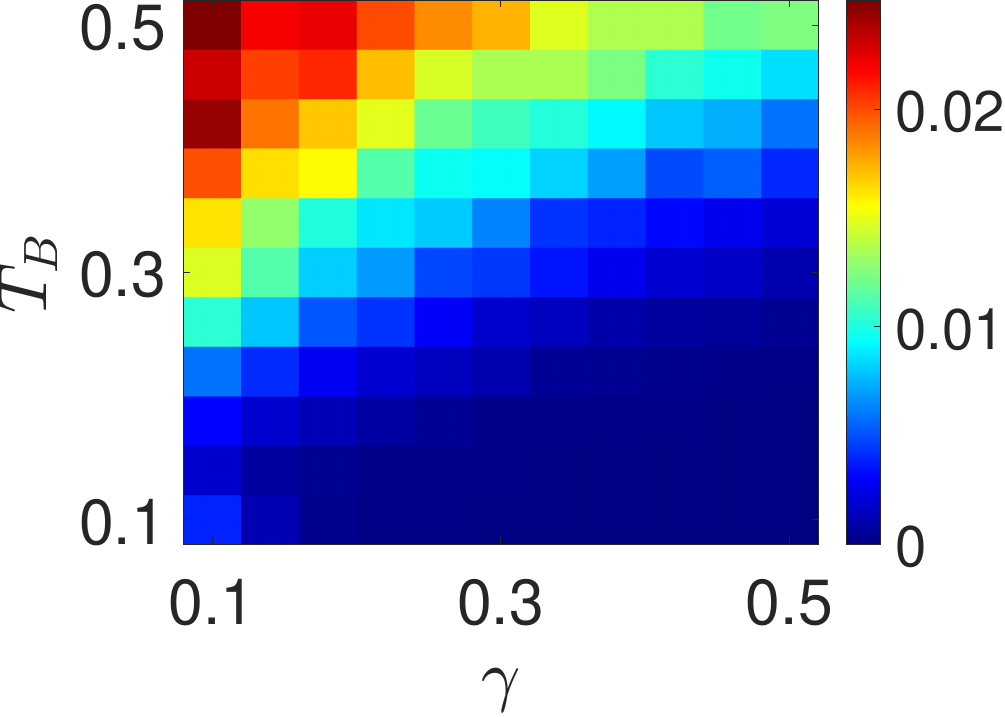}
 	 \caption{}\label{fig:RatesSSEMean}
\end{subfigure}&
\begin{subfigure}[c]{.3\textwidth} 
	  \includegraphics[width=\textwidth]{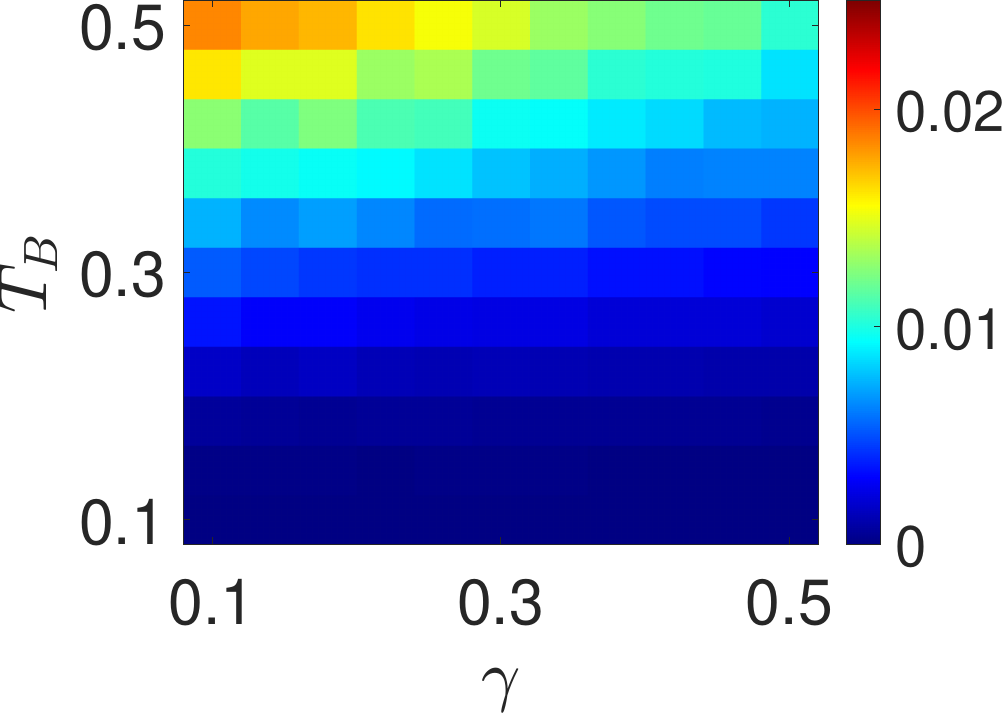}
 	 \caption{}\label{fig:RatesLangevinMean}
\end{subfigure}&
\begin{subfigure}[c]{.3\textwidth} 
	  \includegraphics[width=\textwidth]{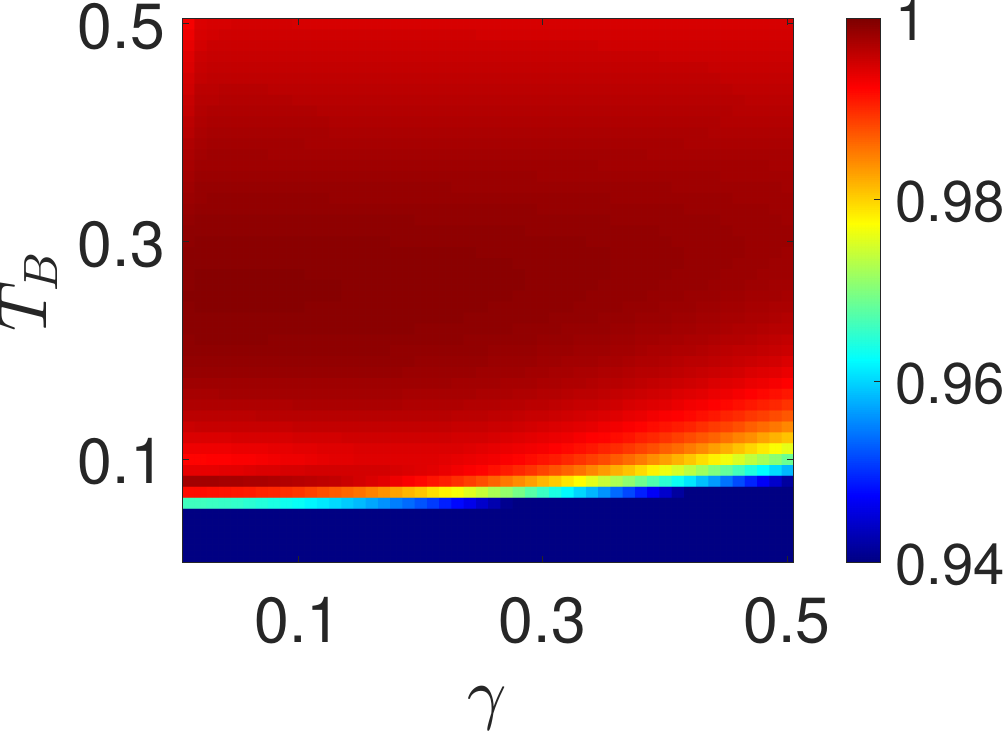}
 	 \caption{}\label{fig:Fidelity}
\end{subfigure}
\end{tabular}
\caption{Left and middle: Transition rates with respect to temperature and coupling strength calculated using Transition Interface Sampling (TIS). The left plot corresponds to SSE rates and the middle plot to Langevin rates. The plot on the right displays the fidelity of the stationary solution of the Lindblad equation with that of the quantum Gibbs state, plotted against temperature and coupling strength. \label{fig:TransitionRate}}
\end{figure*}
\begin{figure}
\centering 
\begin{tabular}{c c}
\begin{subfigure}[c]{.24\textwidth} 
	  \includegraphics[width=\textwidth]{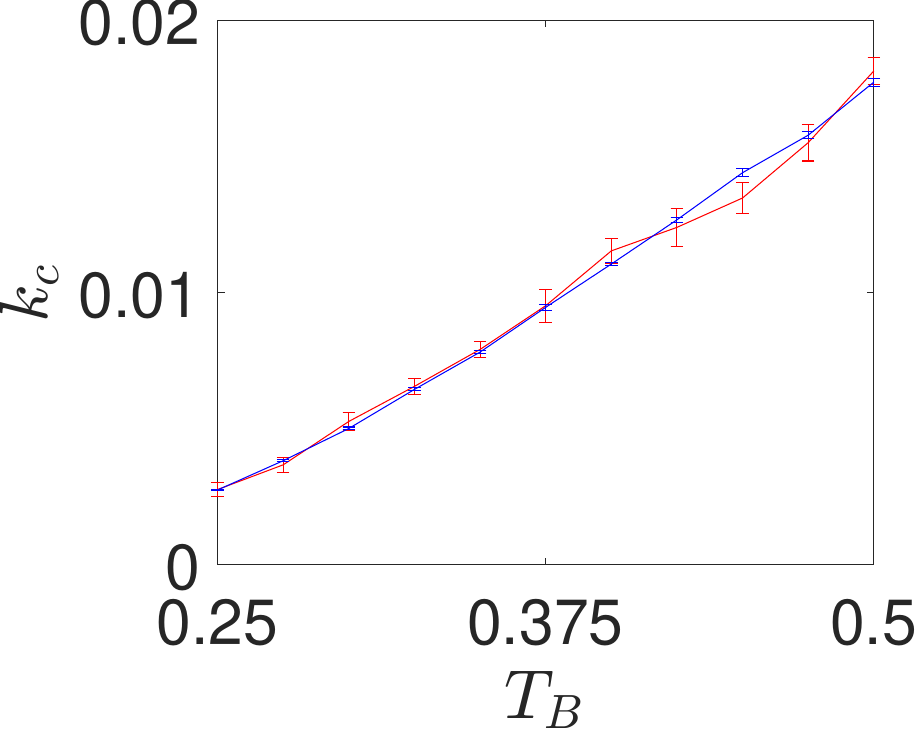}
 	 \caption{}\label{fig:BruteVsTISSSE}
\end{subfigure}&
\begin{subfigure}[c]{.24\textwidth} 
	  \includegraphics[width=\textwidth]{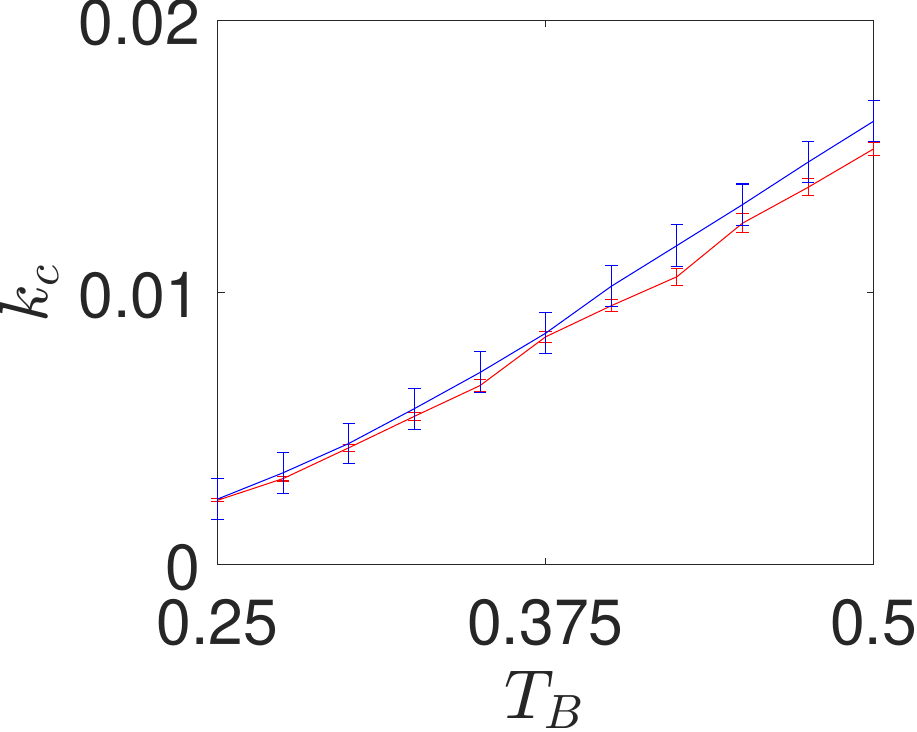}
 	 \caption{}\label{fig:BruteVsTISLangevin}
\end{subfigure}
\end{tabular}
\caption{Comparison of TIS (blue) and MFPT (red) transition rates with respect to temperature at $\gamma=0.25$, the error bars represent the standard error calculated from 10 simulation runs. The left column corresponds to the SSE and the right to the Langevin dynamics. \label{fig:BruteComparison}}
\end{figure}

To study transitions in open quantum systems, we utilize the model of a thermally driven particle in a symmetric quartic double-well potential:
\begin{equation} \label{eq:quartwell}
     V(x)=c_4 x^4-c_2 x^2.
\end{equation}
This model is favored for its simplicity and the presence of metastable states. For simulations, we choose our particle to have the mass of a proton and work in a system with simulation parameters $k_B=1$ and $\hbar=1$. We have standardized the well parameters in all figures as follows:
\begin{equation} \label{eq:WellParam}
c_4=0.01 \quad\text{and}\quad c_2=0.35,
\end{equation}
which yields two minima at $x_{\text{min}}=\pm 4.18$, divided by a barrier with a height of $V_B=3.06$. In SI units, this gives a well-to-well distance of 
\begin{equation}
    L=8.36 \times \sqrt{\frac{\hbar}{m_{p}}}=2.01\times10^{-3} \text{meters},
\end{equation}
and a barrier height of
\begin{equation}
    V_B=k_B \times 3.06=4.22\times 10^{-23} \text{Joules}.
\end{equation}
To offer an intuitive understanding of the temperature's effect relative to the barrier height in our figures, we introduce the dimensionless barrier-normalized temperature, $T_B$, defined by:
\begin{equation}
T_B=\frac{k_B T}{V_B},
\end{equation}
where $k_B$ is the Boltzmann constant, and $T$ is the temperature entering into the SSE and Langevin equation. With this choice of parameters, all rates $k_t$ and friction coefficients $\gamma$ are given in units of inverse seconds ($s^{-1}$).

In \cref{fig:RatesSSEMean,fig:RatesLangevinMean} we display the variation of transition rates with respect to temperature and bath coupling strength calculated using TIS, \cref{fig:RatesSSEMean} contains the SSE transition rates and \cref{fig:RatesLangevinMean} the Langevin rates. In both cases, we observe the typical pattern predicted by Kramers' rate law \cite{mel1991kramers} with the highest rates corresponding to weak coupling and high temperature. The SSE displays overall higher rates with a stronger dependence of the rate on the coupling constant when compared to the Langevin dynamics, we will delve into the mechanisms behind these observations later in this section. \Cref{fig:Fidelity} displays the fidelity~\cite{Fidelity} (a measure of similarity) between the stationary state of the Lindblad equation \eqref{eq:lindblad} and the quantum Gibbs state
\begin{equation}
    \rho_T=e^{-\hat H/k_B T}/\Tr(e^{-\hat H/k_B T}).
\end{equation}
We highlight this because it is critical not to extend the model \eqref{eq:heatbathH} and \eqref{eq:heatbathL} to temperatures too low, as such conditions violate the foundational assumptions of Caldeira and Leggett's master equation derivation~\cite{caldeira1983path}. We observe that for temperatures above \(T_B = 0.1\), the Lindblad stationary state and the quantum Gibbs state closely coincide, suggesting that the Caldeira-Leggett master equation accurately models quantum Brownian motion within this temperature range~\cite{cleary2011phase,christie2024quantum}. To model quantum Brownian motion in this potential at lower temperatures, it becomes crucial to consider non-Markovian bath effects, such as those described by the HPZ master equation~\cite{hu1992quantum,halliwell1996alternative}. While this is an intriguing avenue for future research, it is beyond the scope of the current paper, and we will focus on temperatures above \(T_B = 0.1\).

In \cref{fig:BruteComparison}, we compare transition rates calculated using the inverse mean passage time (iMFPT, depicted in blue) and TIS (depicted in red) within a domain where the application of both methodologies is feasible. Both plots highlight the convergence of the TIS method with standard approaches. A modification to the iMFPT method involves incorporating a cutoff time. This measure addresses the problem of escalating numerical error with increased path length. Specifically, for the iMFPT analysis in this domain, simulations were extended with a cutoff time set at \(t_c = 2000\), translating to a minimum detectable rate of \(k_c = 5 \times 10^{-4}\). The required cutoff time for accurately calculating rates below this threshold via the iMFPT method increases exponentially with decreasing temperature.
\begin{figure*}
\centering 
\begin{tabular}{c c c}
\begin{subfigure}[c]{.3\textwidth} 
	  \includegraphics[width=\textwidth]{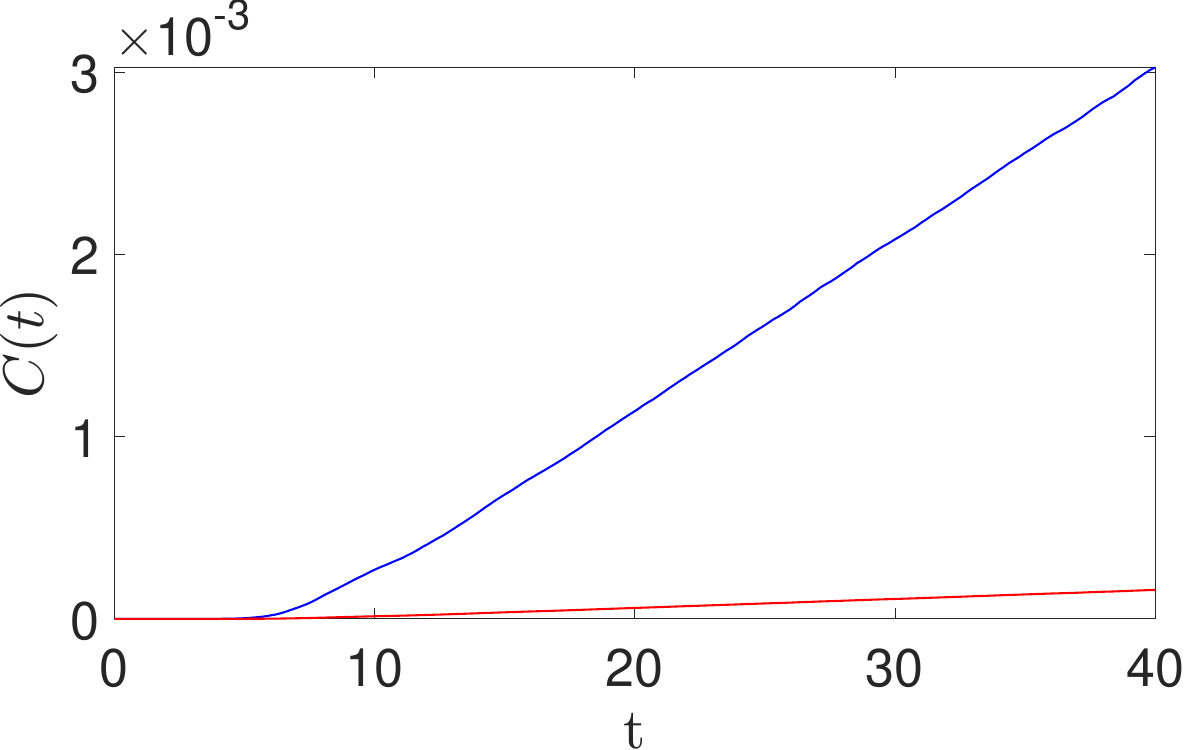}
 	 \caption{}\label{fig:Correlation_SSE_Langevin_T=0.31_Gamma=0.25}
\end{subfigure}&
\begin{subfigure}[c]{.3\textwidth} 
	  \includegraphics[width=\textwidth]{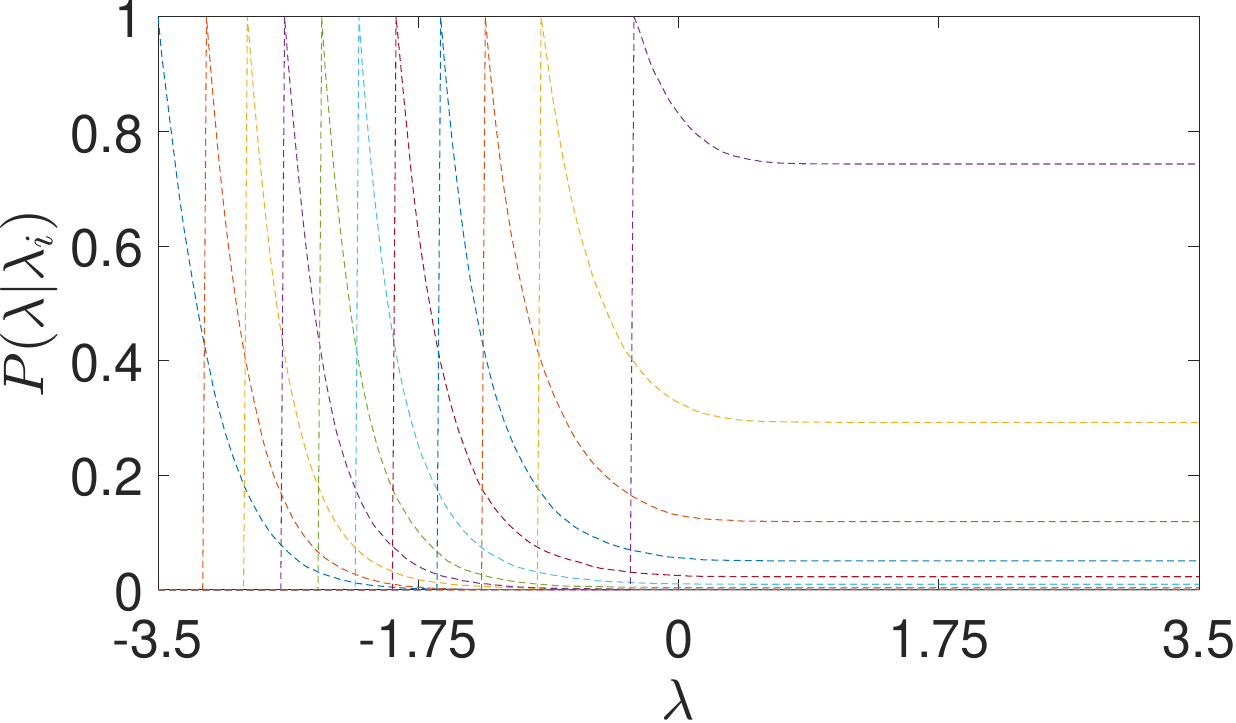}
 	 \caption{}\label{fig:CrossingPlotLangevin}
\end{subfigure}&
\begin{subfigure}[c]{.3\textwidth} 
	  \includegraphics[width=\textwidth]{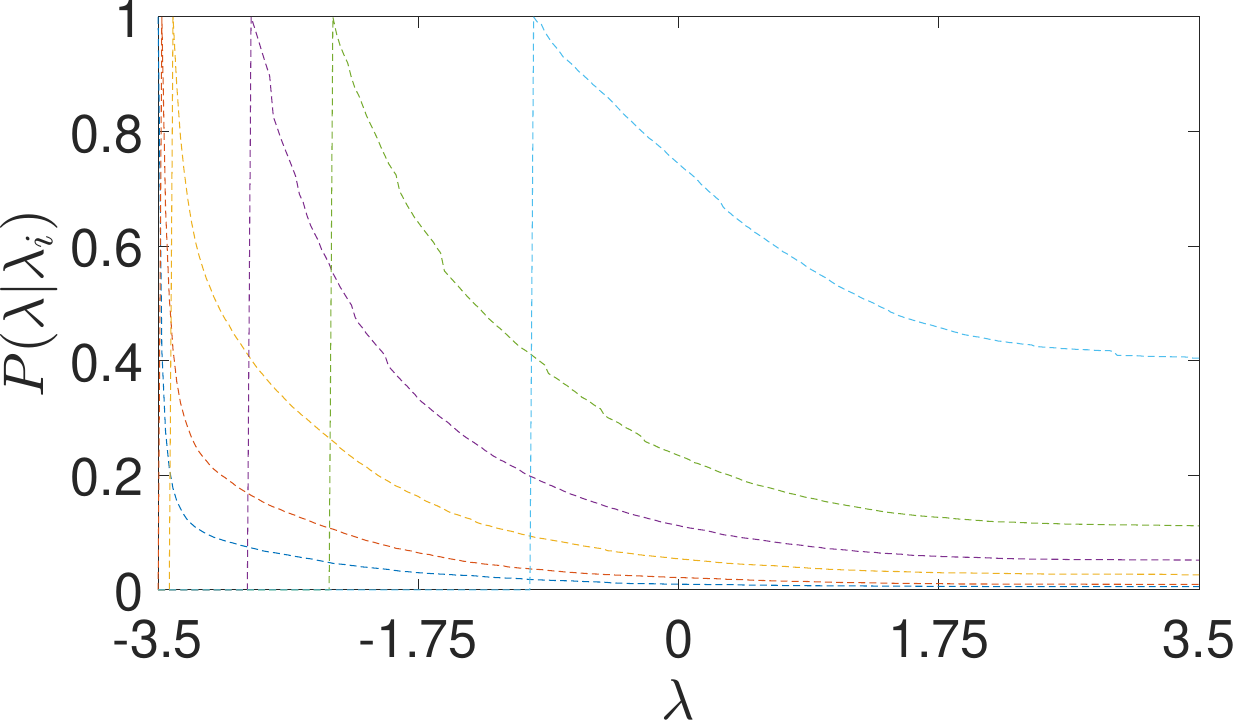}
   \caption{}\label{fig:CrossingPlotSSE}
\end{subfigure}
\end{tabular}
\caption{Left: Langevin (red) and SSE (blue) correlation functions calculated using TPS. The gradient in the regime of linear increase is $k_t=5.9\times10^{-6}$ for the Langevin plot and $k_t=9.3\times10^{-5}$ for the SSE plot. Middle and right Langevin and SSE conditional interface crossing histograms. The TIS rate for the Langevin dynamics is $k_t=5.90\times10^{-6}$ with first interface flux $\Phi_0=4.30\times10^{-3}$. The TIS rate for the SSE dynamics is $k_t=9.30\times10^{-5}$ with first interface flux $\Phi_0=1.30\times10^{-3}$. In all plots, the temperature and coupling strength are given by $T_B=0.1$ and $\gamma=0.25$.  \label{fig:TPS}}
\end{figure*}
\begin{figure*}
\centering 
\begin{tabular}{c c c}
\begin{subfigure}[c]{.3\textwidth} 
	  \includegraphics[width=\textwidth]{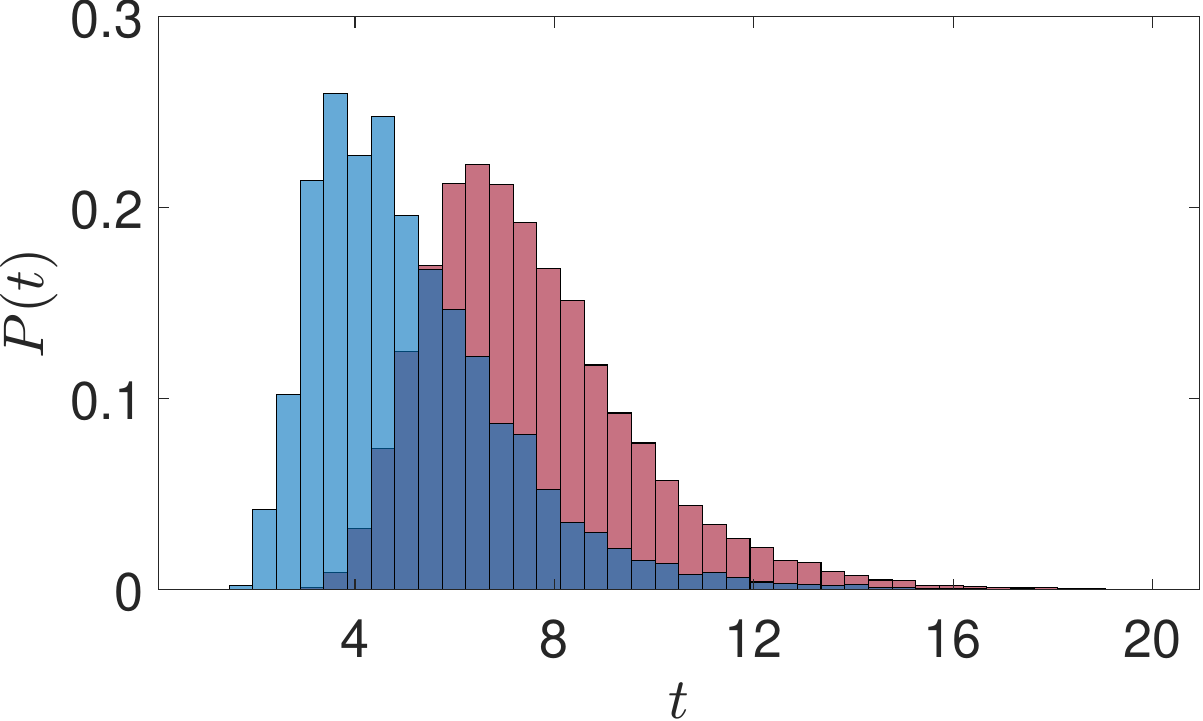}
 	 \caption{}\label{fig:HistPathLength}
\end{subfigure}&
\begin{subfigure}[c]{.3\textwidth} 
	  \includegraphics[width=\textwidth]{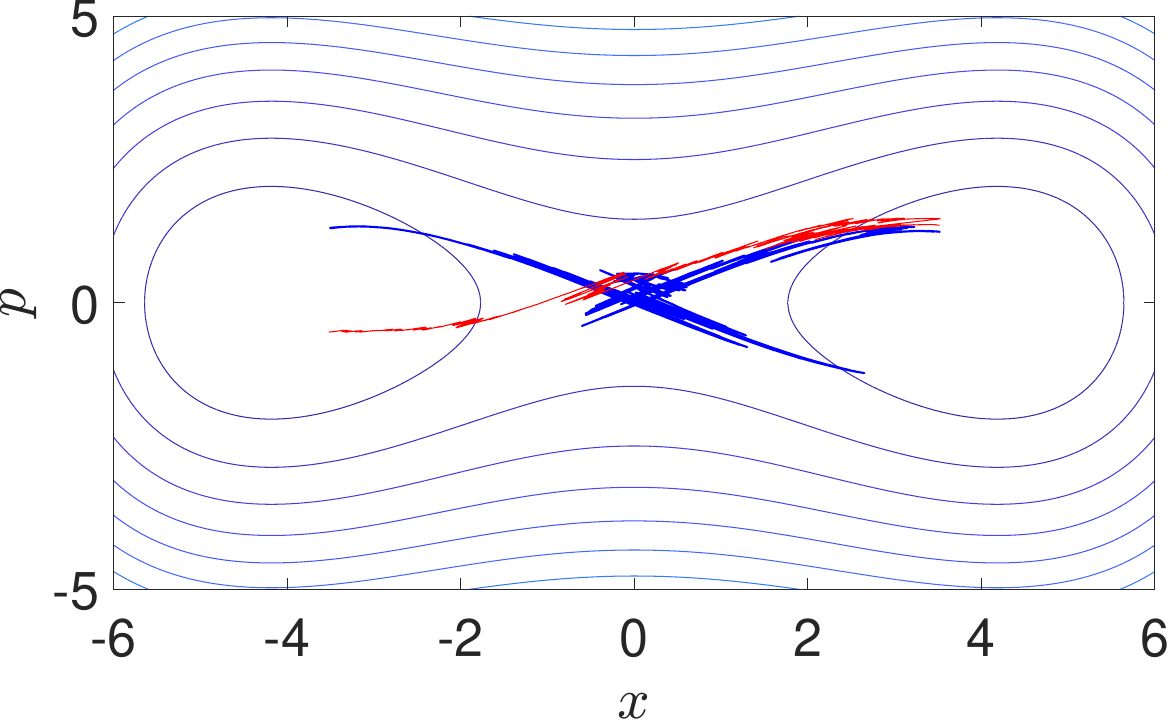}
 	 \caption{}\label{fig:SSEPhaseSingle}
\end{subfigure}&
\begin{subfigure}[c]{.3\textwidth} 
	  \includegraphics[width=\textwidth]{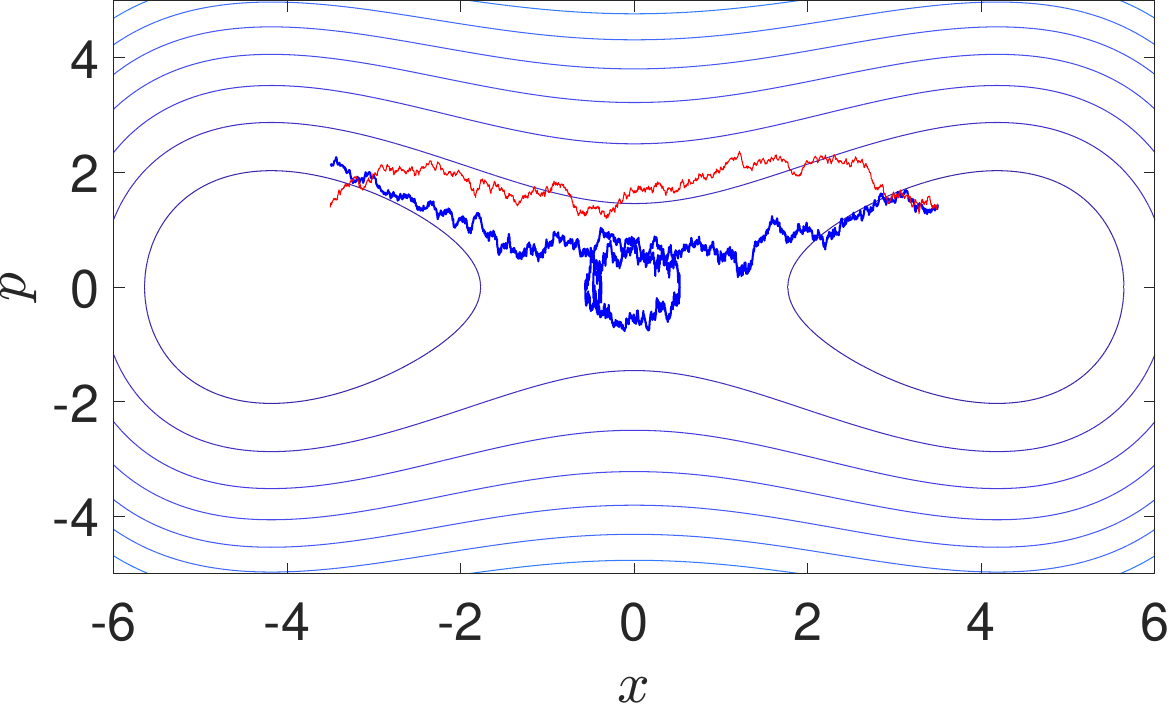}
 	 \caption{}\label{fig:LangevinPhaseSingle}
\end{subfigure}
\end{tabular}
\caption{Left: Probability density histogram of transition path lengths for the SSE (blue) and Langevin dynamics (red) generated using TIS. Middle and right: examples of short (red, $t_f=1.5$) and long (blue, $t_f=15$) trajectories of the SSE (middle) and Langevin (right). The temperature and coupling strength are $T_B=0.1$ and $\gamma=0.25$\label{fig:trajectories}}
\end{figure*}
\begin{figure*}
\centering 
\begin{tabular}{c c c}
\begin{subfigure}[c]{.3\textwidth} 
	  \includegraphics[width=\textwidth]{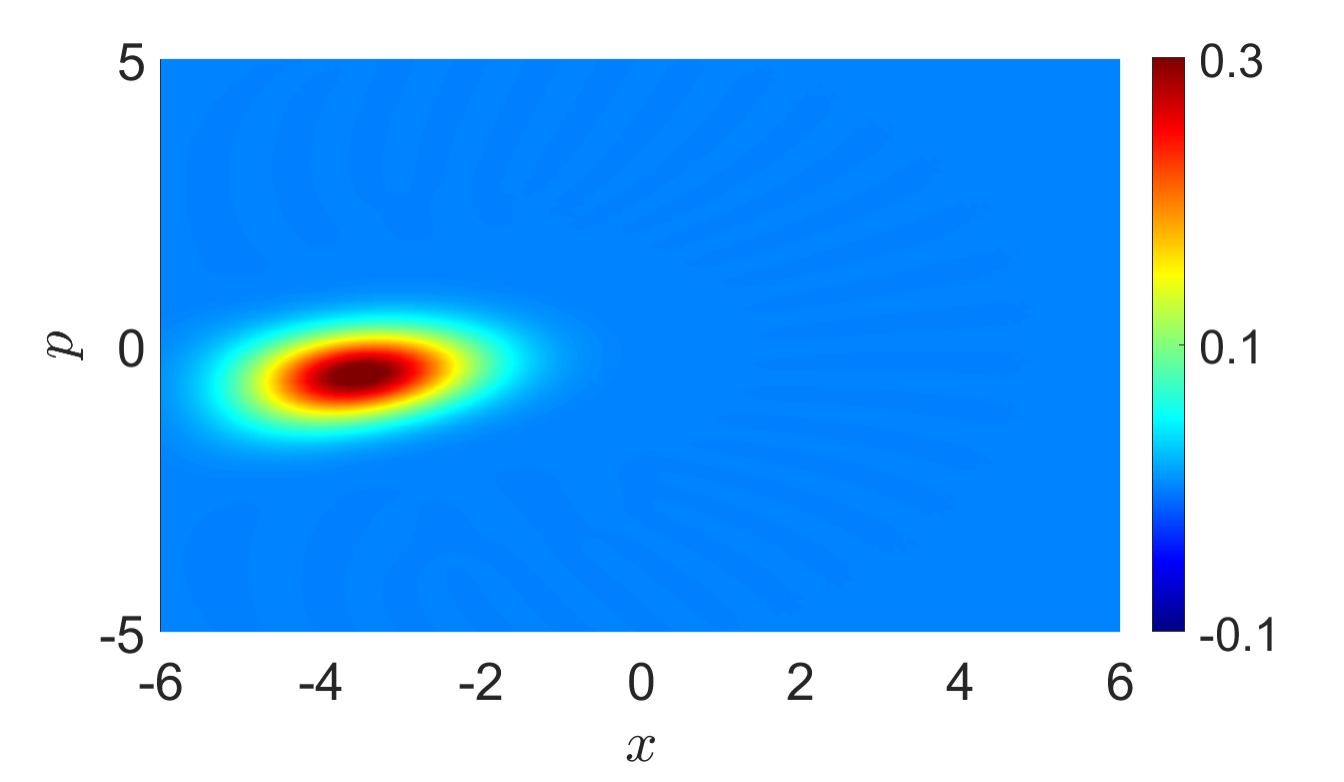}
 	 \caption{$t=0$}\label{fig:WigSnap1}
\end{subfigure}&
\begin{subfigure}[c]{.3\textwidth} 
	  \includegraphics[width=\textwidth]{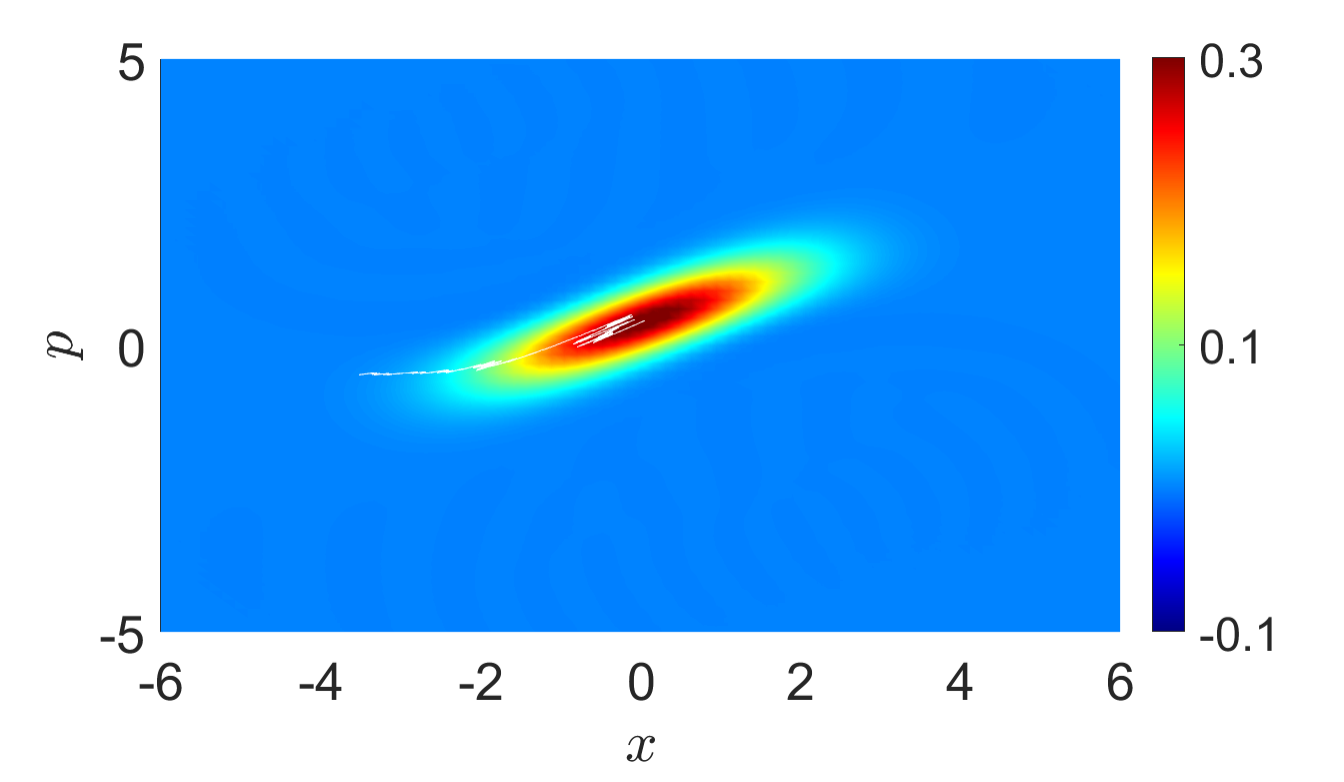}
 	 \caption{$t=0.76$}\label{fig:WigSnap2}
\end{subfigure}&
\begin{subfigure}[c]{.3\textwidth} 
	  \includegraphics[width=\textwidth]{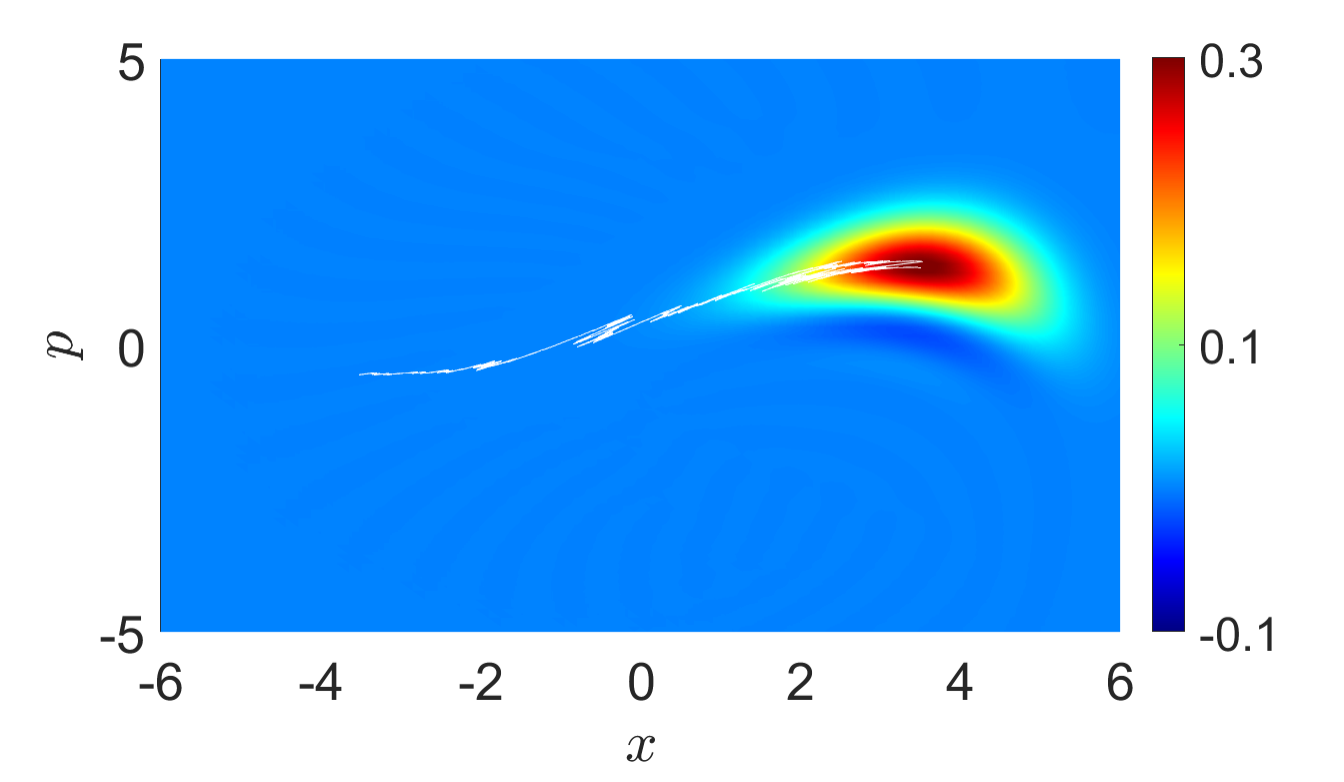}
 	 \caption{$t=1.5$}\label{fig:WigSnap3}
\end{subfigure}\\
\begin{subfigure}[c]{.3\textwidth} 
	  \includegraphics[width=\textwidth]{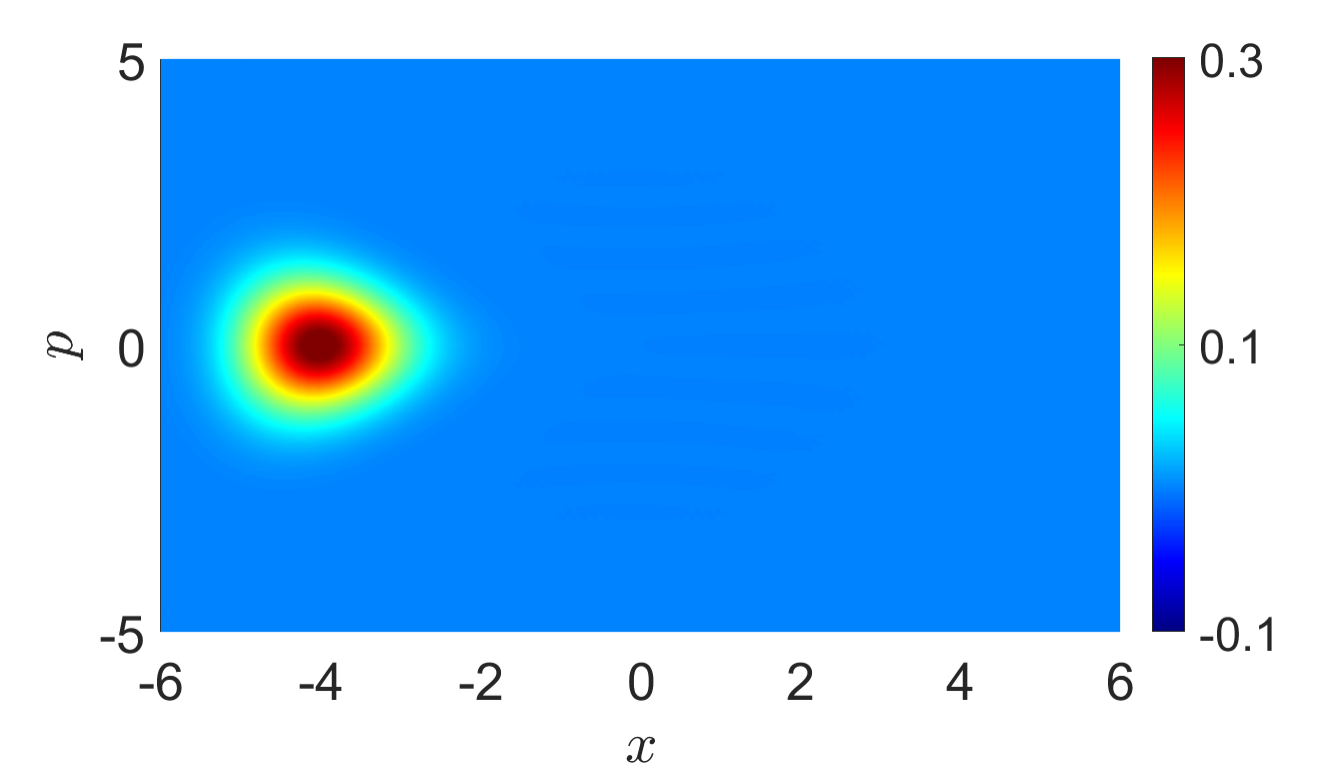}
 	 \caption{$t=0$}\label{fig:WigSnap1Ham}
\end{subfigure}&
\begin{subfigure}[c]{.3\textwidth} 
	  \includegraphics[width=\textwidth]{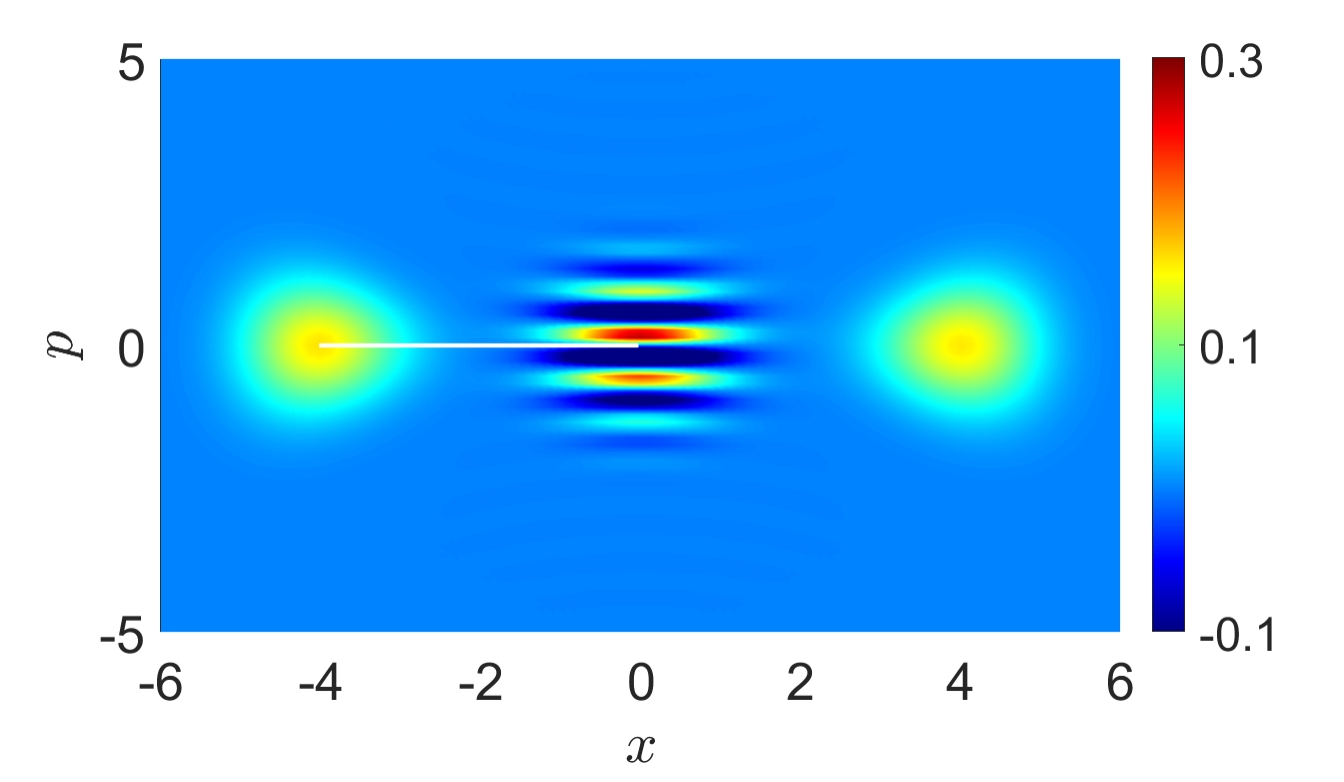}
 	 \caption{$t=1.38\times 10^5$}\label{fig:WigSnap2Ham}
\end{subfigure}&
\begin{subfigure}[c]{.3\textwidth} 
	  \includegraphics[width=\textwidth]{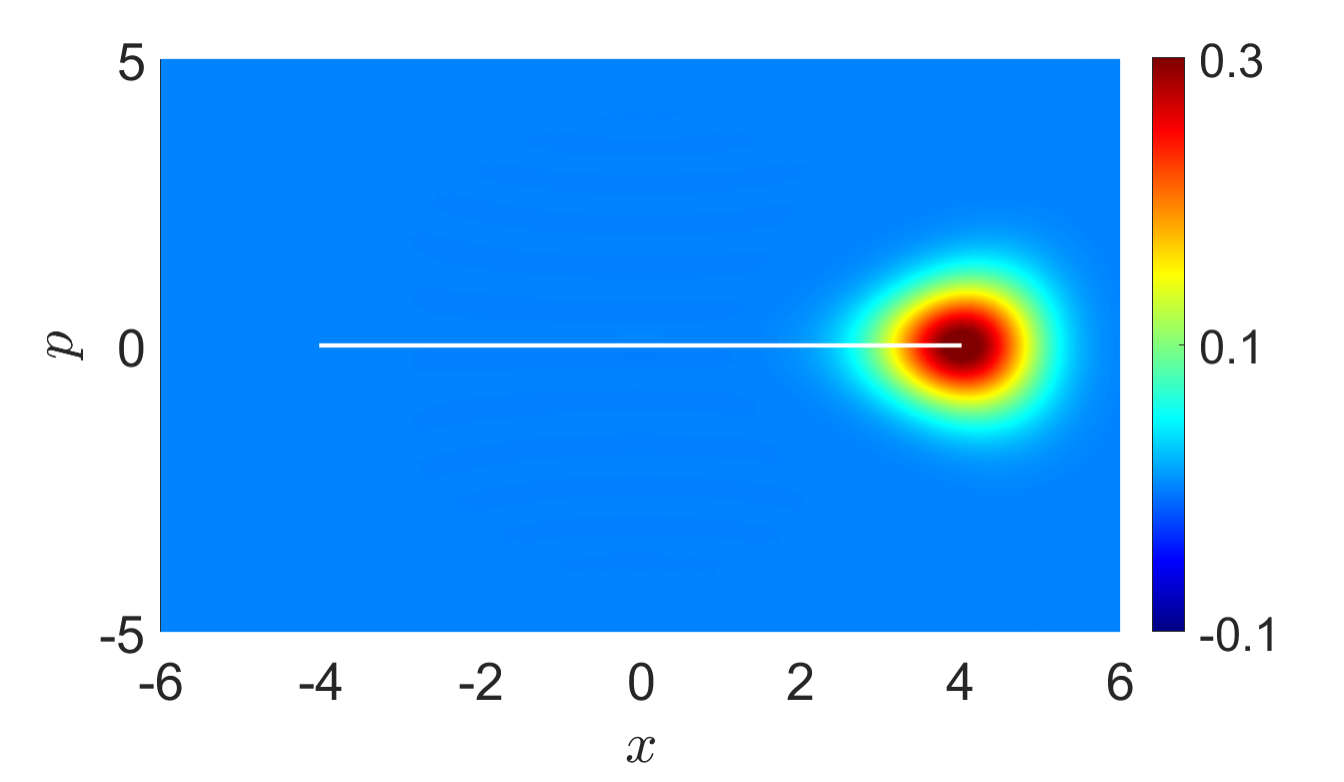}
 	 \caption{$t= 2.76\times 10^5$}\label{fig:WigSnap3Ham}
\end{subfigure}
\end{tabular}
\caption{Top row: Wigner snapshots for the short SSE trajectory in \cref{fig:SSEPhaseSingle} at various times. Bottom row: Wigner snapshots of the coherent tunneling of the decoupled dynamics ($\gamma=0$) for an initially Gaussian state placed at the minima of the left well. The associated rate of this tunnelling is $k_{t}=3.62\times10^{-6}$ \label{fig:WigSnap}}
\end{figure*}

\begin{figure*}
\centering 
\begin{tabular}{c c c}
\begin{subfigure}[c]{.34\textwidth} 
	  \includegraphics[width=\textwidth]{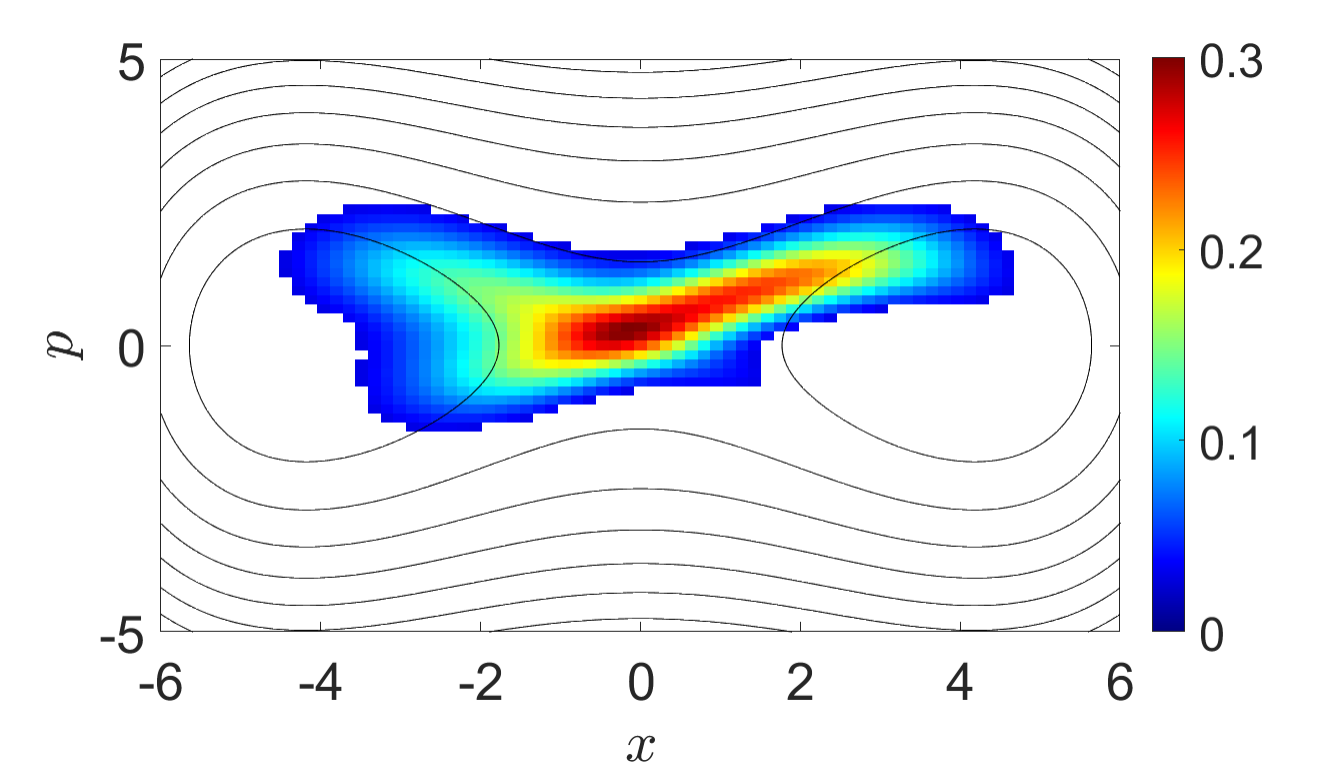}
 	 \caption{\label{fig:SSETransitionHistogramWignerTotal}}
\end{subfigure}&
\begin{subfigure}[c]{.34\textwidth} 
	  \includegraphics[width=\textwidth]{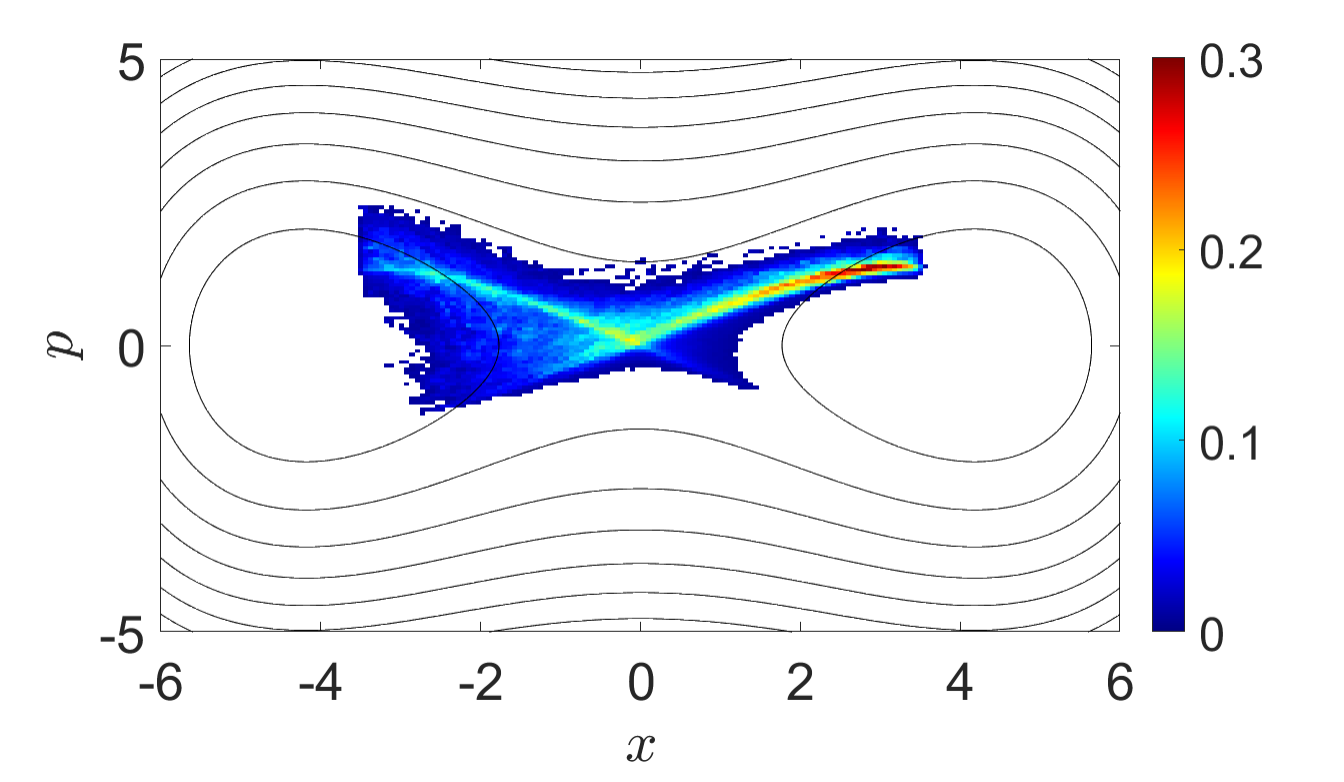}
 	 \caption{}\label{fig:SSETransitionHistogramWignerLow}
\end{subfigure}&
\begin{subfigure}[c]{.34\textwidth} 
	  \includegraphics[width=\textwidth]{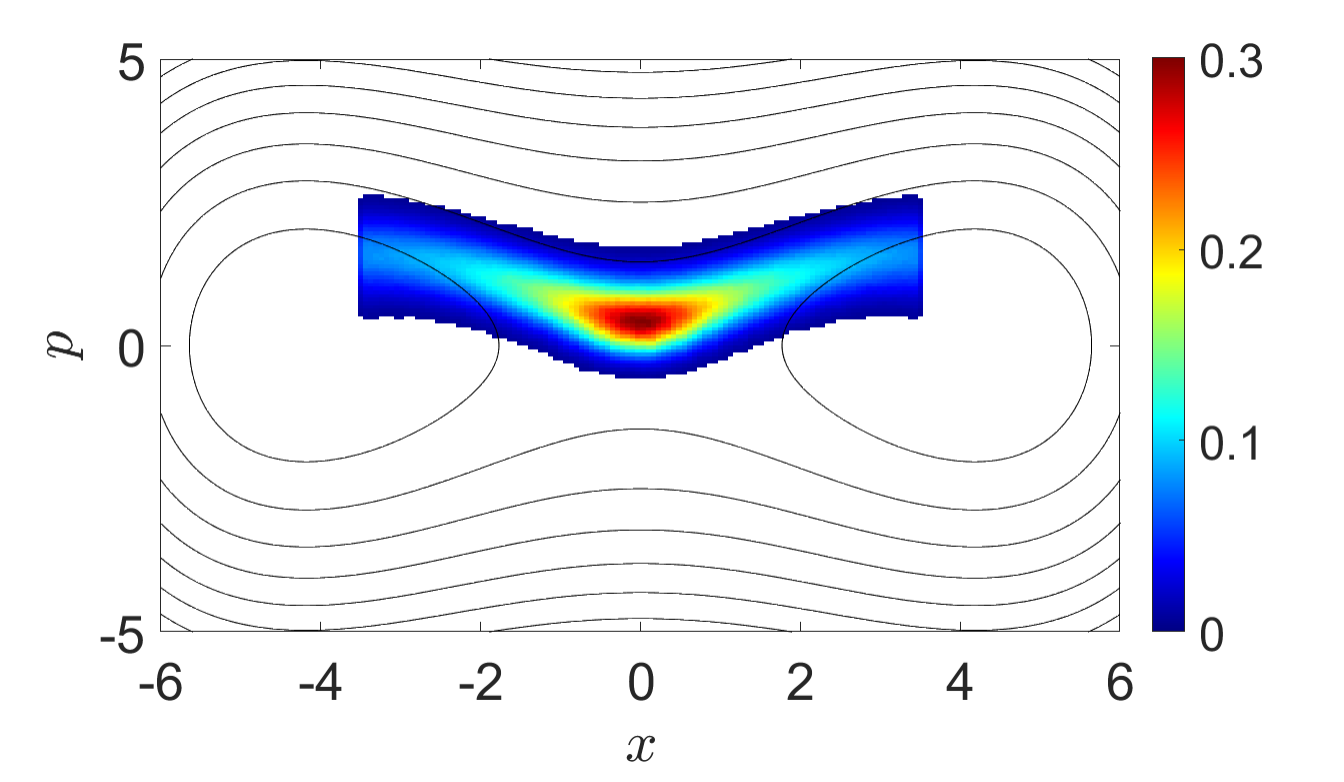}
 	 \caption{}\label{fig:SSETransitionHistogramWignerHigh}
\end{subfigure}
\end{tabular}
\caption{Phase-space heatmaps generated by histogramming: SSE Wigner functions (Left), SSE phase space centers (middle), and Langevin phase space centers (right). All plots are generated using TIS with temperature and coupling strength \(T_B = 0.1\) and \(\gamma = 0.25\). \label{fig:histograms}}
\end{figure*}
\begin{figure}
\centering 
	  \includegraphics[width=0.4\textwidth]{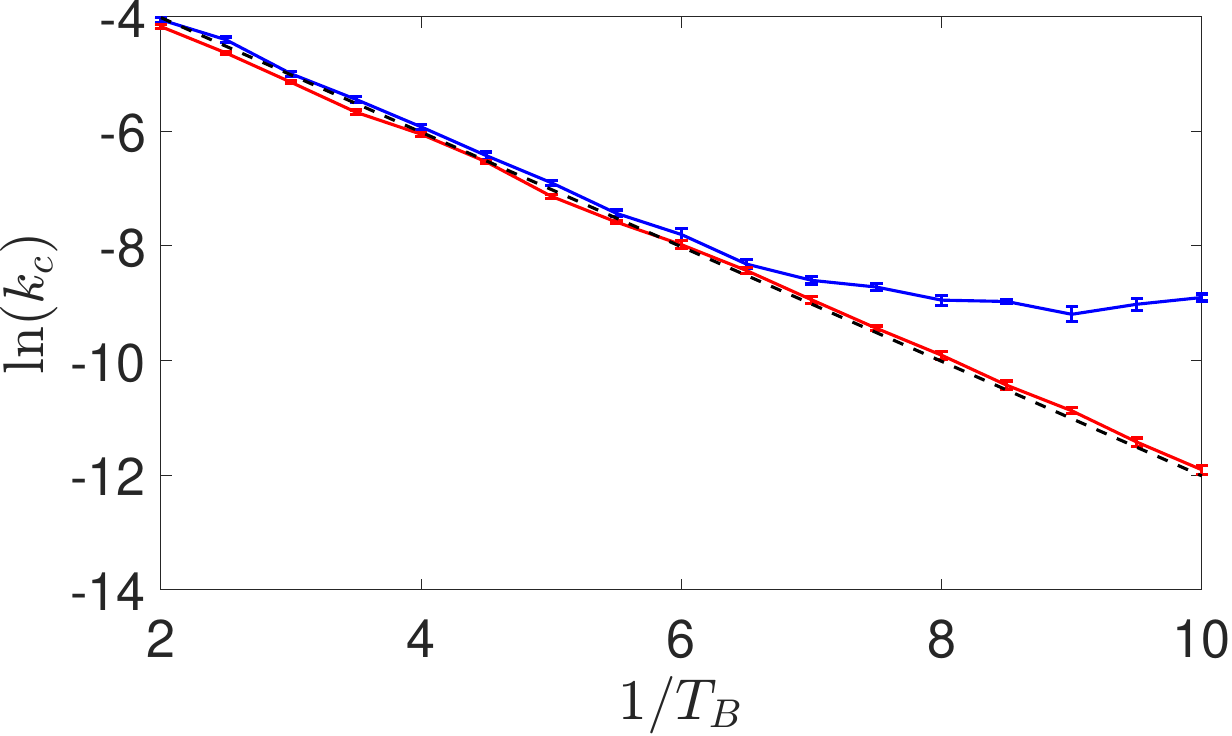}
 	 \caption{\label{fig:ArrheniusPlotSSELine}}
\caption{The logarithm of the transition rate as a function of inverse temperature for the SSE, Langevin, and Arrhenius prediction depicted in blue, red, and black, respectively for the temperature range ($T_B=0.1-0.5$). The error bars represent the standard error of the mean with each point the average of 15 TIS simulations. The coupling strength is given by $\gamma=0.25$\label{fig:Arrhenius}.}
\end{figure}

In \cref{fig:TPS}, we show the results of TPS and TIS simulations at a bath temperature of $T_B=0.1$ and a damping coefficient of $\gamma=0.25$. \cref{fig:Correlation_SSE_Langevin_T=0.31_Gamma=0.25} presents the TPS-calculated correlation functions \eqref{eq:Correl} for both Langevin and SSE dynamics. We observed that the SSE correlation function exhibits a steeper gradient in the linear regime, with $k_t \approx 9.3 \times 10^{-5}$, compared to $k_t \approx 5.9 \times 10^{-6}$ for the Langevin correlation. The use of the approximate sign ($\approx$) is due to some ambiguity in defining the precise start of the linear regime. \Cref{fig:CrossingPlotLangevin,fig:CrossingPlotSSE} display the conditional crossing probabilities \eqref{eq:TISRate} for the Langevin and SSE TIS simulations, respectively. The interfaces were dynamically placed to ensure a probability of $0.4$ for reaching successive interfaces at all stages except the final interface, following \cite{borrero2011optimizing}. The Langevin crossing histograms are typically more evenly spaced due to inertial effects preventing fast recrossings. We found good agreement between the TPS and TIS-calculated rates but with significantly reduced resource requirements in the TIS case. A key difference between Transition Path Sampling (TPS) and Transition Interface Sampling (TIS) that is absent in classical models is the need for more basis states in TPS for accurate simulations. This requirement arises because, in TPS, trajectories extend into regions A and B and continue without terminating even after reaching stable states A or B. Consequently, this results in an expanded configuration space that necessitates a larger basis for precise representation.

A histogram of the TIS path lengths for the SSE and Langevin dynamics is presented in \cref{fig:HistPathLength}, illustrating that the Langevin transition paths are, on average, longer than those of SSE. In \cref{fig:SSEPhaseSingle,fig:LangevinPhaseSingle}, the phase space centers' trajectories are shown, including examples of short (red) and long (blue) trajectories for both SSE and Langevin dynamics. As anticipated, short Langevin trajectories are associated with higher average momentum than long Langevin trajectories. This observation is further illustrated in the right plot of the histograms shown in Figure \ref{fig:histograms}. Notably, the short SSE trajectory is observed to start with a negative momentum expectation value and approach the separatrix with a negative momentum expectation value before transitioning. Snapshots of the Wigner function for this short SSE trajectory are shown in the top row of \cref{fig:WigSnap}. We observe that this trajectory remains predominantly Gaussian throughout the transition, with minor interference fringing most evident as the center crosses the separatrix. Given the near-Gaussian nature of this state throughout the transition, we can heuristically explain this puzzling behavior using a Gaussian approximation to the SSE dynamics. This is accomplished by assuming the state remains Gaussian and then computing the semiclassical (up to order $\hbar$) expansion of the SSE dynamics about the phase space center. Detailed formulations of these types of calculations can be found in \cite{MyPaper,myThesis}, leading us to the following expression for the centroid dynamics:
\begin{multline} \label{eq:Gaussian_Double}
    	\begin{pmatrix}
    	    d\ev*{\hat{X}} \\ d\ev*{\hat{P}}
    	\end{pmatrix}=\left[\frac{1}{m}\begin{pmatrix} \ev*{\hat{P}}\\0 \end{pmatrix}-2\gamma \ev*{\hat{P}}\begin{pmatrix} 0\\1\end{pmatrix}- \begin{pmatrix} 0\\V'(\ev*{\hat{X}}) \end{pmatrix} \right] dt\\
     +  \sqrt{4 \gamma m k_B T \left(1 - \frac{4 \Delta \hat{P}^2 \Delta \hat{X}^2}{\hbar^2}\right) }\begin{pmatrix}0\\1\end{pmatrix} d \xi \\
     -\left( \sqrt{\frac{\hbar^2 \gamma}{4 m  k_B T}} - \frac{4 \Delta \hat{X}^2 \sqrt{m k_B T \gamma}}{\hbar}\right)\begin{pmatrix}1 \\ 0\end{pmatrix}
 d \xi
\end{multline}
here, $\ev*{\hat{X}}$ and $\ev*{\hat{P}}$ represent the position and momentum expectation values, respectively, while $\Delta \hat X$ and $\Delta \hat P$ denote the associated variances. A notable point of contrast between the Gaussian SSE \eqref{eq:Gaussian_Double} and the Langevin equation \eqref{eq:Langevin_tilde} is that in the Gaussian SSE, the stochastic term can directly influence the position expectation value. This mechanism is a form of the anti-Zeno effect \cite{AntiZeno1,AntiZeno2,maniscalco2006zeno}, where the bath may be thought of as repeatedly measuring the position of the quantum particle. Despite the outcomes of these measurements being random, in rare instances, these measurements repeatedly localize the particle by shifting it in the same direction, providing a momentum-free transition mechanism not present in the classical case. We found that the vast majority of SSE transition trajectories start in the positive momentum region, and those transitions with strictly negative momentum expectation values up to the separatrix are exceedingly rare and difficult to observe without path sampling methods. We stress that the observed anti-Zeno effect differs from coherent tunneling since states restricted to Gaussian forms cannot tunnel. For comparison, in the bottom row of \cref{fig:WigSnap}, we show the coherent tunneling of an initially Gaussian state in the same potential but disconnected from the bath ($\gamma=0$). We observe a vast difference in timescales between the anti-Zeno and coherent tunneling transitions, with the tunneling transition taking on the form of a cat-state \cite{castanos2012dynamics} at the separatrix and the anti-Zeno effect remaining mostly Gaussian with minor interference fringing. A key observation to be gained here is that while the coupling of a quantum particle to a heat bath suppresses tunneling, it can also induce an anti-Zeno mechanism. Revisiting Figure~\ref{fig:TPS}, we conclude that at a temperature \( T_B = 0.1 \), this anti-Zeno mechanism accounts for the dominant contribution to the overall transition rate of \( k_t = 9.30 \times 10^{-5} \). If we assume that the thermal barrier hopping rate is given by the rate calculated via Langevin dynamics, \( k_t = 5.9 \times 10^{-6} \), and the unsuppressed (\( \gamma = 0 \)) tunneling rate in this potential is \( k_t = 3.62 \times 10^{-6} \) (see Figure~\ref{fig:WigSnap}).

Figure \ref{fig:histograms} examines the phase-space. From left to right we display phase-space heatmaps from histogrammed Wigner functions for the SSE, histogrammed phase space centers for the SSE and histogrammed Langevin phase space centers. We note differences between the SSE and Langevin cases: an asymmetry between the left and right wells in the SSE case is attributed to the time irreversibility of SSE dynamics. An `X' shape is evident in the SSE histograms, resulting from the previously mentioned anti-Zeno transition paths.

Figure \ref{fig:Arrhenius} presents transition rates for the SSE, Langevin, and Arrhenius-predicted dynamics. The Arrhenius prediction was found by fitting a line with a gradient of $-V_{B}$ to the Langevin data in \cref{fig:ArrheniusPlotSSELine} using least squares error minimization. The Langevin rates align with the Arrhenius predictions at low temperatures, while the quantum rates diverge from the simple functional form of Arrhenius law, likely due to additional anti-Zeno transition mechanisms. The quantum rate in this figure also displays good agreement with the analytically derived rate law in Ref. \cite{topaler1994quantum}.

\section{Summary}
In this study, we applied transition path sampling to trajectories formed by stochastic Schrödinger dynamics in Markovian open quantum systems driven by Gaussian noise. Our focus was on a model rooted in a Caldeira-Leggett oscillator bath, but the methodology is applicable to all systems described by stochastic Schrödinger equations.

The TIS and TPS simulations conducted revealed a wealth of information on the phase-space dynamics of the Langevin equation and SSE. We observed slightly longer SSE transition paths compared to Langevin dynamics and noted a significant asymmetry in the transition path histograms in the SSE case due to irreversible dynamics. We find strong alignment in Langevin rates at low temperatures with the Arrhenius law and a notable divergence in quantum rates from Arrhenius predictions due to an additional anti-Zeno transition mechanism.

Much in the same way that path sampling with underlying Langevin dynamics has been utilized to uncover classical reaction mechanisms, increases in computing power now facilitate further exploration of quantum molecular dynamics when combined with rare-event methods. This exploration holds the potential to uncover mechanisms of low-temperature reactions that would otherwise be classically forbidden.

\section{Acknowledgements}
R. Christie acknowledges support from EPSRC grant no.EP/W522673/1 and computing support provided by the Imperial College Research Computing Service \cite{imperial_college_2023}. R. Christie is currently located at the University of Portsmouth, UK.

\section{Code Availability}
The MATLAB code used to generate the data contained in this paper is available on GitHub at 
\href{https://github.com/rchristie95/SSEPathSampling}{https://github.com/rchristie95/SSEPathSampling}.

\section*{References}
\bibliographystyle{IEEEtran}
\bibliography{main}{}
\appendix
\section{Path probability of stochastic Schr\"{o}dinger trajectories} \label{sec:SSE-Derivation}
We calculate the short-time propagator for the $d$ component complex stochastic process with a single discrete Euler step, so
\begin{equation} \label{eq:SSEReal}
    \psi''=\psi'+u\left(\psi'\right) \delta t+\sigma\left(\psi'\right) \delta W,
\end{equation}
where $\delta W$ normally distributed with a variance of $\delta t$. The drift and diffusion coefficients are given by
\begin{equation}
    u\left(\psi\right)=\frac{1}{\hbar}\left(-{i}\hat{H} -\frac{1}{2 }\hat{L}^{\dagger}\hat{L}+\ev*{\hat{L}^{\dagger}}\hat{L} - \frac{1}{2}\ev*{\hat{L}^{\dagger}}\ev*{\hat{L}}\right)\ket {\psi},
\end{equation}
and 
\begin{equation}
    \sigma\left(\psi\right)=\frac{1}{\sqrt{ \hbar}}\left(\hat{L}-\ev*{\hat{L}}\right)\ket{\psi}.
\end{equation}
To ensure all later integrals are well defined, we map this to a $2 d$ dimensional real representation, with operators $\hat A$ and complex number $i$ mapped to
\begin{equation}
    A_{\mathbb{C}}\to A_{\mathbb{R}}=\begin{pmatrix}
        \Re(A)&-\Im(A)\\\Im(A)&\Re(A)
    \end{pmatrix},
\end{equation}
and
\begin{equation}
i\to J=\begin{pmatrix}
        0&-I_d\\ I_d&0
    \end{pmatrix},
\end{equation}
respectively. Since $\delta W$ is Gaussian distributed, the transition probability distribution from $\psi'$ to $\psi''$ is given by
\begin{equation} \label{eq:Noise_Hist}
    \mathcal{P}\left(\psi'_{0} \to \psi''_{\delta t}\right)=\frac{1}{\sqrt{2 \pi \delta t}} \exp(-\frac{\delta W^2}{2 \delta t}).
\end{equation}
To write the above in terms of states $\psi''$ and $\psi'$ we introduce a delta function and integrate over the dummy noise variable $\delta \tilde{W}$ as follows
\begin{multline}
    \mathcal{P}\left(\psi'_{0} \to \psi''_{\delta t}\right)=\\\int_{-\infty}^{\infty}\mathrm{d} \delta \tilde{W} \delta\left(\psi''-\left(\psi'+u(\psi') \delta t+ \sigma(\psi') \delta \tilde{W}\right)\right)  \mathcal{P}\left(\delta \tilde{W}\right).
\end{multline}
We can insert a Fourier representation of the delta function, which gives
\begin{multline}
    \mathcal{P}\left(\psi'_{0} \to \psi''_{\delta t}\right)=\frac{1}{(2 \pi)^{2d}} \int \mathrm{d}^{2d} \boldsymbol{k} \int \mathrm{d} \delta \tilde{W} \\e^{i \boldsymbol{k} \cdot\left(\psi''-\left(\psi'+u(\psi') \delta t+ \sigma(\psi') \delta \tilde{W}\right)\right)}  \mathcal{P}\left(\delta \tilde{W}\right).
\end{multline}
Performing the integral over the $\delta \tilde{W}$ variable gives
\begin{multline} \label{eq:fourier}
\mathcal{P}\left(\psi'_{0} \to \psi''_{\delta t}\right)  =\\\frac{1}{(2 \pi)^{2 d-\frac12}\sqrt{\delta t}} \int \mathrm{d}^{2d} \boldsymbol{k} e^{i \boldsymbol{k} \cdot\left(\psi''-\left(\psi'+u(\psi') \delta t\right)\right)-\frac{1}{2} \delta t(\boldsymbol{k} \cdot \mathbf{g}(\psi') \cdot \boldsymbol{k})}
\end{multline}
with
\begin{equation}
    \mathbf{g}(\psi(t)=\sigma(\psi)\sigma(\psi)^{T}.
\end{equation}
The usual procedure for integrating $\boldsymbol{k}$ assumes that $\mathbf{g}(\psi(t)$ is invertible (full rank) and in the results in the path probability\cite{wio2013path}
\begin{equation} 
\mathcal{P}[\Psi] \propto e^{-\frac{1}{2} \int_0^{t_f}(\dot{\Psi}(t)-u(\Psi(t))) \cdot \mathbf{g}(\Psi(t))^{-1} \cdot(\dot{\Psi}(t)-u(\Psi(t))) \mathrm{d} t}.
\end{equation}
However in our case $\mathbf{g}(\Psi(t)$ is the projection operator 
\begin{equation}
\mathbf{g} =\frac{1}{\hbar}\left(L\left(\psi \psi^T\right) L^T-\langle L\rangle\left(L \psi \psi^T+\psi \psi^{T} L^{T}\right)+\langle L\rangle^2 \psi \psi^T\right)
\end{equation}
and for $\boldsymbol{k}\in\text{null}(\boldsymbol{g})$ the integral \eqref{eq:fourier} becomes infinite. If we want the Fourier integral to be well defined, we must instead take the $\boldsymbol{k}$ integral over only the rank space of $\mathbf{g}(\psi(t)$ which in the simple case of real noise we can add this constraint in the form of a delta function as
\begin{equation}
    \frac{1}{(2 \pi)^{2d}} \int \mathrm{d}  \lambda  \int \mathrm{d}^{2d} \boldsymbol{k} f(\boldsymbol{k})\delta(\boldsymbol{k}-\lambda \sigma(\psi)) = \frac{1}{(2 \pi)^{2d}} \int \mathrm{d} \lambda f(\lambda \sigma(\psi)).
\end{equation}
At first glance, this seems a bit sketchy; however, we are only integrating over increments that can be generated by the SSE (however improbable) after a single timestep. In contrast to the invertible diffusion matrix case, not all increments $\boldsymbol{k}\in \mathbb{R}^{2 d}$ may be generated in a single time step, with the SSE generating only multiples of the noise term of the form
\begin{equation}
    \boldsymbol{k}=\lambda \sigma(\psi)\quad \text{for}\quad \lambda \in \mathbb{R}.
\end{equation}
The transition probability for a single timestep is thus
\begin{multline}
    \mathcal{P}\left(\psi'_{0} \to \psi''_{\delta t}\right)=\frac{1}{(2 \pi)^{2d-\frac12}\sqrt{\delta t}} \int \mathrm{d}\lambda \\  e^{i \lambda \sigma(\psi') \cdot\left(\psi''-\left(\psi'+u(\psi') \delta t\right)\right)-\frac{\lambda^2}{2} \delta t(\sigma(\psi') \cdot \sigma(\psi'))^2}.
\end{multline}
Performing the $\lambda$ integral yields the step probability
\begin{multline}
    \mathcal{P}\left(\psi'_{0} \to \psi''_{\delta t}\right)=\frac{1}{(2 \pi)^{2d-1}\sigma(\psi')\cdot\sigma(\psi')\delta t}\\\exp{-\frac{ \left(\sigma(\psi')\cdot\left(\psi''-\psi'- u(\psi')\delta t\right) \right)^2}{2 \delta t(\sigma(\psi')\cdot\sigma(\psi'))^2}}.
\end{multline}
Comparing the above with our initial expression for the transition probability \eqref{eq:Noise_Hist} we can see that they must have the same prefactor and we may set the above prefactor as follows
\begin{multline}\label{eq:P_Step}
    \mathcal{P}\left(\psi'_{0} \to \psi''_{\delta t}\right)=\\\frac{1}{\sqrt{2 \pi \delta t}}\exp{-\frac{ \left(\sigma(\psi')\cdot\left(\psi''-\psi'- u(\psi')\delta t\right) \right)^2}{2 \delta t(\sigma(\psi')\cdot\sigma(\psi'))^2}}.
\end{multline}
We can convert back to the complex representation at this stage and obtain the one-step probability
\begin{multline}\label{eq:P_StepC}
    \mathcal{P}\left(\psi'_{0} \to \psi''_{\delta t}\right)=\\\frac{1}{\sqrt{2 \pi \delta t}}\exp{-\frac{ \left|\sigma^{\dagger}(\psi')\left(\psi''-\psi'- u(\psi')\delta t\right) \right|^2}{2 \delta t(\sigma^{\dagger}(\psi')\sigma(\psi'))^2}}.
\end{multline}
If we are renormalizing at each time step following \cref{alg:Euler-NSSE} we simply make the replacement
\begin{equation}
    \psi'\to \frac{\psi'}{\norm{\psi'}}
\end{equation}
in \eqref{eq:P_StepC}. For a trajectory $\Psi(t)$ in the continuum limit we have
\begin{equation} \label{eq:SSEActionRealCRep}
P[\Psi]\! \propto\! \exp \left\{\!-\! \int_0^{t_f}\mathrm{d} t\frac{\left|\sigma^{\dagger}(\Psi(t))\dot{\Psi}(t)-\sigma^{\dagger}(\Psi(t)) u(\Psi(t))\right|^2 }{2(\sigma^{\dagger}(\Psi(t))\sigma(\Psi(t)))^2} \right\}.
\end{equation}
We can carry out a similar derivation for the one-step probability amplitude
\begin{equation}
A\left(\psi'_{0} \to \psi''_{\delta t}\right)=\sqrt{\mathcal{P}\left(\psi'_{0} \to \psi''_{\delta t}\right)}e^{i \theta}.
\end{equation}
Here $\theta$ is the phase change accumulated during the Euler step and is given by the expression
\begin{equation}\theta=\arctan\left(\frac{\Im\left(1+u^{\dagger}\psi' \delta t+\sigma^{\dagger}\psi' \delta W\right)}{\Re\left(1+u^{\dagger}\psi' \delta t+\sigma^{\dagger}\psi' \delta W\right)}\right).
\end{equation}
For the SSE we have
\begin{equation}
    \sigma(\psi')=\frac{1}{\sqrt{ \hbar}}(\hat{L}-\ev*{\hat{L}}_{\psi'}) \ket {\psi'},
\end{equation}
and thus
\begin{equation}
    \sigma^{\dagger}(\psi') \psi'=0,
\end{equation}
Therefore
\begin{equation}
    \theta=\arctan\left(\frac{- i \left(u^{\dagger}(\psi')\psi'-u(\psi')\psi'^{\dagger}\right) \delta t}{2+\left(u^{\dagger}\psi'+u(\psi')\psi'^{\dagger}\right) \delta t}\right).
\end{equation}
For small $\delta t$ we have
\begin{equation}
    \theta\approx \frac{- i}{2} \left(u^{\dagger}(\psi')\psi'-u(\psi')\psi'^{\dagger}\right) \delta t=\Im \left(u^{\dagger}(\psi')\psi'\right) \delta t.
\end{equation}
Thus we can write for the continuum limit amplitude
\begin{multline} \label{eq:SSEAmplituteRealCRep}
A[\Psi] \propto \exp \bigg\{\int_0^{t_f}\bigg(-\frac{\left|\sigma^{\dagger}(\psi(t))\dot{\psi}(t)-\sigma^{\dagger}(\psi(t)) u(\psi(t))\right|^2 }{4(\sigma^{\dagger}(\psi(t))\sigma(\psi(t)))^2}\\+i \Im \left(u^{\dagger}(\psi(t))\psi(t)\right) \bigg)\mathrm{d} t\bigg\}.
\end{multline}We are not restricted to a single Gaussian noise source, an alternative form of the SSE (also known as quantum state diffusion) is of the form \cite{percival}
\begin{equation} \label{eq:SSEComplex}
    \psi''=\psi'+u\left(\psi'\right) \delta t+\frac{1}{\sqrt{2}}\sigma\left(\psi'\right)( \delta W^R+ i \delta W^I).
\end{equation}
The situation is similar to the real noise case, with the exception that the diffusion tensor will now be the rank-2 projector
\begin{equation}
    \boldsymbol{g}=\frac12(\sigma+J \sigma)(\sigma+J \sigma)^T=\sigma \sigma^T+J\sigma \sigma^T-\sigma \sigma^T J- J\sigma \sigma^T J.
\end{equation}
we can follow a similar derivation as above and obtain the expression for the path action for a multi-segment path with $\delta t\to0$ as 
\begin{multline} 
\mathcal{P}[\Psi] \propto \exp \bigg\{\\- \int_0^{t_f}\mathrm{d} t\frac{\bigg((I_{2d}+J)\sigma(\Psi(t))\cdot\dot{\Psi}(t)
-(I_{2d}+J)\sigma(\Psi(t))\cdot u(\Psi(t)\bigg)^2 }{2(\sigma(\Psi(t)\cdot\sigma(\Psi(t)))^2)} \bigg\},
\end{multline}
or in complex representation
\begin{equation}
    \mathcal{P}[\Psi] \propto \exp \bigg\{- \int_0^{t_f}\mathrm{d} t\frac{\abs{\sigma^{\dagger}(\Psi(t))\cdot\dot{\Psi}(t)-\sigma^{\dagger}(\Psi(t))\cdot u(\Psi(t)}^2 }{4(\sigma^{\dagger}(\Psi(t)\cdot\sigma(\Psi(t)))^2)} \bigg\}.
\end{equation} 
\section{Involutory Transformations and Mirror TPS} \label{app:MirrorTPS}
\begin{figure}
\begin{tikzpicture}[scale=0.7, decoration={markings, mark=at position 0.2 with {\arrow{>}}, mark=at position 0.4 with {\arrow{>}}, mark=at position 0.7 with {\arrow{>}}, mark=at position 0.9 with {\arrow{>}}}]
\draw[thick,->] (-2,0,0) -- (8,0,0) node[right]{$t$};
\draw[thick,->] (0,-2,0) -- (0,2,0) node[above]{$p$};
\draw[thick,->] (0,0,-2) -- (0,0,2) node[below left]{$q$};
\foreach \i/\j in {0/$t_0$ , 1.5/, 3/$t_s$ , 4.5/, 6/$t_N$} {
    \fill[red, opacity=0.3] (\i,2,-2) -- (\i,2,-1) -- (\i,-2,-1) -- (\i,-2,-2) -- cycle;
    \fill[gray, opacity=0.1] (\i,2,-1) -- (\i,2,1) -- (\i,-2,1) -- (\i,-2,-1) -- cycle;
    \fill[blue, opacity=0.3] (\i,2,1) -- (\i,2,2) -- (\i,-2,2) -- (\i,-2,1) -- cycle;
    \ifx\j\empty\else
        \node at (\i-0.5, -3, 0) {\j};
    \fi
}
\draw[postaction={decorate}, violet, thick, domain=0:6, samples=10, smooth] plot (\x, {cos(deg(0.5*\x))^2}, {-1.3*cos(deg(0.5*\x))}) node[ right]{$\Psi^{o}(t)$};
\foreach \i in {1.5,3,4.5} {
    \fill[violet] (\i, {cos(deg(0.5*\i))^2}, {-1.3*cos(deg(0.5*\i))}) circle (1.5pt);
}
\draw [fill=white] (0, {cos(deg(0.5*0))^2}, {-1.3*cos(deg(0.5*0))}) circle [radius=.07];
\fill[black] (6, {cos(deg(0.5*6))^2}, {-1.3*cos(deg(0.5*6))}) circle (1.5pt);
\draw[postaction={decorate}, red, thick, domain=0:6, samples=25, smooth] plot (\x, {-1+2*cos(deg(0.5*\x))^2}, {0.0464437+1.60147*cos(deg(0.545211*\x))}) node[right]{$\Psi^{n}(t)$};
\foreach \i in {1.5,3,4.5} {
    \fill[red] (\i, {-1+2*cos(deg(0.5*\i))^2}, {0.0464437+1.60147*cos(deg(0.545211*\i))}) circle (1.5pt);
}
\draw [fill=white] (0, {cos(deg(0.5*0))^2}, {0.0464437+1.60147*cos(deg(0.545211*0))}) circle [radius=.07];
\fill[black] (6, {-1+2*cos(deg(0.5*6))^2}, {0.0464437+1.60147*cos(deg(0.545211*6))}) circle (1.5pt);
\draw[postaction={decorate}, blue, thick, domain=0:6, samples=10, smooth] plot (\x, {1.2*sin(deg(0.5*\x))^2}, {-1.7*cos(deg(0.5*\x))}) node[ right]{$\mathcal{S}\Psi^{n}(t)$};
\foreach \i in {1.5,3,4.5} {
    \fill[blue] (\i, {1.2*sin(0.5*deg(\i))^2}, {-1.7*cos(deg(0.5*\i))}) circle (1.5pt);
}
\draw [fill=white] (0, {1.2*sin(0.5*deg(0))^2}, {-1.7*cos(deg(0.5*0))}) circle [radius=.07];
\fill[black] (6, {1.2*sin(0.5*deg(6))^2}, {-1.7*cos(deg(0.5*6))}) circle (1.5pt);
\draw[black, <-, thick] 
  (3, {-1+2*cos(deg(0.5*3))^2+0.1}, {-1.3*cos(deg(0.5*3))}) -- 
  (3, {cos(deg(0.5*3))^2}, {-1.3*cos(deg(0.5*3))})
    node[pos=0.1, right, font=\scriptsize] {$\Delta p$};
\end{tikzpicture}
    \caption{Two-way shooting with path adjustment: The trajectory $\Psi^{n}$ (red) is generated by forward and backward shooting from the original trajectory $\Psi^{o}$ (purple) at $t=t_s$. If $\Psi^{n}$ fails to be reactive but the transformation $\mathcal{S}$ maps $\Psi^{n}$ to a reactive trajectory we may accept $\mathcal{S}\Psi^{n}$ (blue) with probability law \eqref{eq:probMultiShot}.}
    \label{fig:2-Way-Shooting-Sym}
\end{figure}
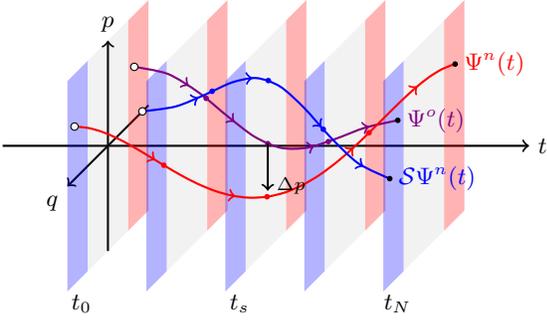
\begin{figure}
    \centering
    \begin{tikzpicture}[scale=0.4]
\tikzmath{
  function sinecurve(\x) {
    return {5+sin(\x * 180 / pi)};
  };
  function negsinecurve(\x) {
    return {-sin(\x * 180 / pi)};
  };
    function coscurve(\x) {
    return {5+cos(\x * 180 / pi)};
  };
  function negcoscurve(\x) {
    return {-cos(\x * 180 / pi)};
  };
}
\draw[<->, >=stealth, line width=0.5mm] (0.7*pi,1) -- (0.7*pi,4) 
        node[midway, right] {$S$};

\draw[<->, >=stealth, line width=0.5mm] (4*pi,1) -- (4*pi,4) 
        node[midway, right] {$S$};

\draw[red, domain=0:2*pi, smooth, variable=\x, thick] plot ({\x},{sinecurve(\x)});
\node[red, below] at (3*pi/2,{sinecurve(3*pi/2)}) {$t_s$};
\fill[red] (3*pi/2,{sinecurve(3*pi/2)}) circle (4pt);
\node[red, right] at (2*pi,{sinecurve(2*pi)}) {$\Psi^{(\mathrm{o})}$};
\draw[red, dashed, bend left=65, decoration={markings,
mark=at position 0.5 with {\arrow[scale=3]{>}}},
postaction={decorate}] (3*pi/2,{sinecurve(3*pi/2)}) to (9*pi/2,{coscurve(3*pi/2)});
\draw[blue, domain=0:2*pi, smooth, variable=\x, thick] plot ({\x},{negsinecurve(\x)});
\node[blue, above] at (pi/2,{negsinecurve(pi/2)}) {$t_{s'}$};
\fill[blue] (pi/2,{negsinecurve(pi/2)}) circle (4pt);
\node[blue, right] at (2*pi,{negsinecurve(2*pi)}) {$S\Psi^{(\mathrm{o})}$};
\draw[blue, dashed, bend right=65, decoration={markings,
mark=at position 0.5 with {\arrow[scale=3]{<}}},
postaction={decorate}] (pi/2,{negsinecurve(pi/2)}) to (7*pi/2,{negcoscurve(7*pi/2)});
\draw[red, domain=3*pi:5*pi, smooth, variable=\x, thick] plot ({\x},{coscurve(\x)});
\node[red, below] at (9*pi/2,{coscurve(3*pi/2)}) {$t_s$};
\fill[red] (9*pi/2,{coscurve(3*pi/2)}) circle (4pt);
\node[red, right] at (5*pi,{coscurve(5*pi)}) {$\Psi^{(\mathrm{n})}$};

\draw[blue, domain=3*pi:5*pi, smooth, variable=\x, thick] plot ({\x},{negcoscurve(\x)});
\node[blue, above right] at (7*pi/2,{negcoscurve(pi/2)}) {$t_{s'}$};
\fill[blue] (7*pi/2,{negcoscurve(pi/2)}) circle (4pt);
\node[blue, right] at (5*pi,{negcoscurve(5*pi)}) {$S\Psi^{(\mathrm{n})}$};
\end{tikzpicture}
    \caption{Schematic of mirror TPS with one transformation operator $S$. Depicted are the trajectories $\Psi^{(\mathrm{o})}$, $S\Psi^{(\mathrm{o})}$, $\Psi^{(\mathrm{n})}$ and $S\Psi^{(\mathrm{n})}$ corresponding to the old, transformed-old, new and transformed-new trajectories. As the transformation may alter the shooting time we make the distinction between $t_s$ and $t_{s'}=S(t_s)$.}
    \label{fig:MultiShotSym}
\end{figure}
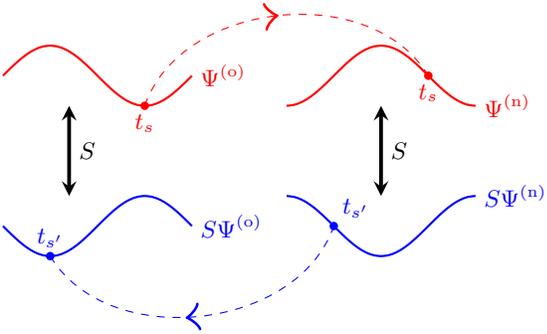
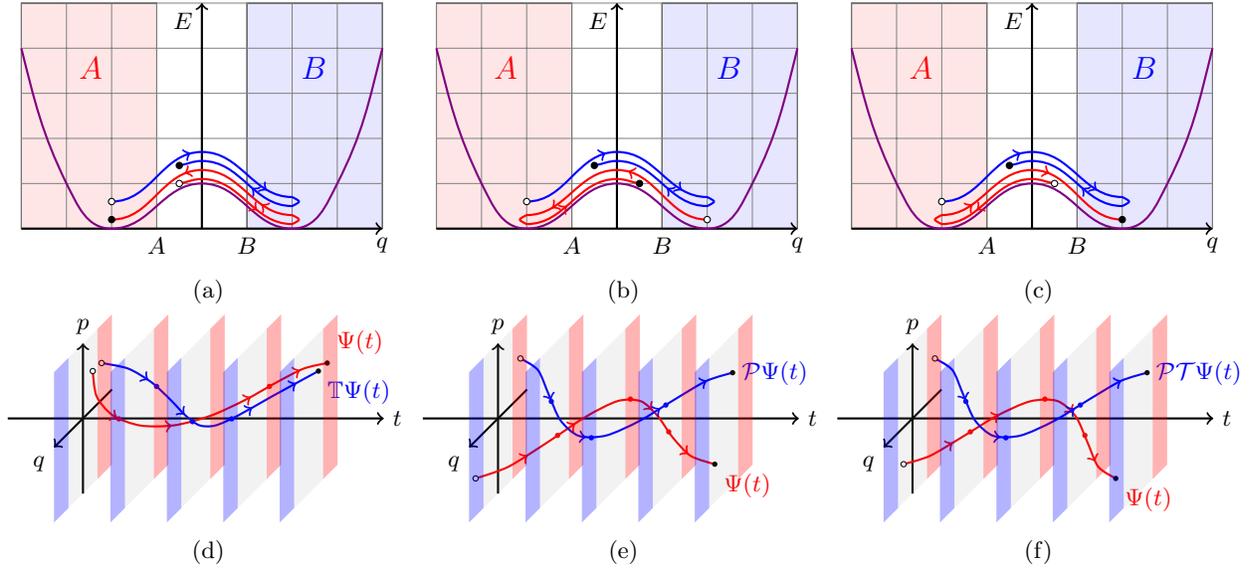
\begin{figure*}
\centering
  \begin{tabular}[c]{c c c}
\begin{subfigure}[c]{.3\textwidth}
\centering
\begin{tikzpicture}[scale=0.6,
    decoration1/.style={decoration={markings,mark=at position 0.25 with {\arrow{>}},mark=at position 0.5 with {\arrow{>}},mark=at position 0.75 with {\arrow{>}}}},
    decoration2/.style={decoration={markings,mark=at position 0.25 with {\arrow{<}},mark=at position 0.5 with {\arrow{<},},mark=at position 0.75 with {\arrow{<}}}}
]
\draw [fill=blue!10] (1,0) rectangle (4,5) ;
\draw [fill=red!10] (-4,0) rectangle (-1,5) ;
\draw[help lines] (-4,0) grid (4,5) ;
\draw [->, thick]  (-4,0) --(-1,0) node[below]{$A$}--(1,0)node [below] {$B$}-- (4,0) node[below]{$q$} ;
\draw [->, thick]   (0,0) -- (0,5) node[below left]{$E$} ;
\draw [thick, violet] (-4,4)  to [out=284,in=116]  (-3,1) to [out=297,in=180]  (-2,0) to [out=0,in=180]  (0,1) to [out=0,in=180]    (2,0) to [out=0,in=243]   (3,1) to [out=63,in=256] (4,4) ;
\draw [postaction={decorate}, decoration1, thick, red] (-0.5,1) to [out=10,in=180]  (0,1.1) to [out=0,in=180] (2,0.1) .. controls (2.2,0.2) and (2.2,0.2) .. (2,0.3) to [out=180,in=0] (0,1.3) to [out=180,in=0] (-2,0.2)  ;
\draw [postaction={decorate}, decoration2, thick, blue] (-0.5,1.4) to [out=10,in=180]  (0,1.5) to [out=0,in=180] (2,0.5) .. controls (2.2,0.6) and (2.2,0.6) .. (2,0.7) to [out=180,in=0] (0,1.7) to [out=180,in=0] (-2,0.6);
\draw [fill=black] (-0.5,1.4) circle [radius=.07];
\draw [fill=white] (-2,0.6) circle [radius=.07];
\draw [fill=white] (-0.5,1) circle [radius=.07];
\draw [fill=black] (-2,0.2) circle [radius=.07];
\draw [thick, red](-2,4)node[below left]{\large{$A$} } ;
\draw [thick, blue](2,4)node[below right]{\large{$B$} } ;
\end{tikzpicture}
\caption{\label{fig:T1}}
\end{subfigure}&
\begin{subfigure}[c]{.3\textwidth}
\centering
\begin{tikzpicture}[scale=0.6,
    decoration1/.style={decoration={markings,mark=at position 0.25 with {\arrow{>}},mark=at position 0.5 with {\arrow{>}},mark=at position 0.75 with {\arrow{>}}}},
    decoration2/.style={decoration={markings,mark=at position 0.25 with {\arrow{<}},mark=at position 0.5 with {\arrow{<}},mark=at position 0.75 with {\arrow{<}}}}
]
\draw [fill=blue!10] (1,0) rectangle (4,5) ;
\draw [fill=red!10] (-4,0) rectangle (-1,5) ;
\draw[help lines] (-4,0) grid (4,5) ;
\draw [->, thick]  (-4,0) --(-1,0) node[below]{$A$}--(1,0)node [below] {$B$}-- (4,0) node[below]{$q$} ;
\draw [->, thick]   (0,0) -- (0,5) node[below left]{$E$} ;
\draw [thick, violet] (-4,4)  to [out=284,in=116]  (-3,1) to [out=297,in=180]  (-2,0) to [out=0,in=180]  (0,1) to [out=0,in=180]    (2,0) to [out=0,in=243]   (3,1) to [out=63,in=256] (4,4) ;
\draw [postaction={decorate}, decoration2, thick, red, xscale=-1] (-0.5,1) to [out=10,in=180]  (0,1.1) to [out=0,in=180] (2,0.1) .. controls (2.2,0.2) and (2.2,0.2) .. (2,0.3) to [out=180,in=0] (0,1.3) to [out=180,in=0] (-2,0.2)  ;
\draw [postaction={decorate}, decoration2, thick, blue] (-0.5,1.4) to [out=10,in=180]  (0,1.5) to [out=0,in=180] (2,0.5) .. controls (2.2,0.6) and (2.2,0.6) .. (2,0.7) to [out=180,in=0] (0,1.7) to [out=180,in=0] (-2,0.6);
\draw [fill=black] (-0.5,1.4) circle [radius=.07];
\draw [fill=white] (-2,0.6) circle [radius=.07];
\draw [fill=black] (0.5,1) circle [radius=.07];
\draw [fill=white] (2,0.2) circle [radius=.07];
\draw [thick, red](-2,4)node[below left]{\large{$A$} } ;
\draw [thick, blue](2,4)node[below right]{\large{$B$} } ;
\end{tikzpicture}
\caption{\label{fig:P1}}
\end{subfigure}&
\begin{subfigure}[c]{.3\textwidth}
\centering
\begin{tikzpicture}[scale=0.6,
    decoration1/.style={decoration={markings,mark=at position 0.25 with {\arrow{>}},mark=at position 0.5 with {\arrow{>}},mark=at position 0.75 with {\arrow{>}}}},
    decoration2/.style={decoration={markings,mark=at position 0.25 with {\arrow{<}},mark=at position 0.5 with {\arrow{<}},mark=at position 0.75 with {\arrow{<}}}}
]
\draw [fill=blue!10] (1,0) rectangle (4,5) ;
\draw [fill=red!10] (-4,0) rectangle (-1,5) ;
\draw[help lines] (-4,0) grid (4,5) ;
\draw [->, thick]  (-4,0) --(-1,0) node[below]{$A$}--(1,0)node [below] {$B$}-- (4,0) node[below]{$q$} ;
\draw [->, thick]   (0,0) -- (0,5) node[below left]{$E$} ;
\draw [thick, violet] (-4,4)  to [out=284,in=116]  (-3,1) to [out=297,in=180]  (-2,0) to [out=0,in=180]  (0,1) to [out=0,in=180]    (2,0) to [out=0,in=243]   (3,1) to [out=63,in=256] (4,4) ;
\draw [postaction={decorate}, decoration1, thick, red, xscale=-1] (-0.5,1) to [out=10,in=180]  (0,1.1) to [out=0,in=180] (2,0.1) .. controls (2.2,0.2) and (2.2,0.2) .. (2,0.3) to [out=180,in=0] (0,1.3) to [out=180,in=0] (-2,0.2)  ;
\draw [postaction={decorate}, decoration2, thick, blue] (-0.5,1.4) to [out=10,in=180]  (0,1.5) to [out=0,in=180] (2,0.5) .. controls (2.2,0.6) and (2.2,0.6) .. (2,0.7) to [out=180,in=0] (0,1.7) to [out=180,in=0] (-2,0.6);
\draw [fill=black] (-0.5,1.4) circle [radius=.07];
\draw [fill=white] (-2,0.6) circle [radius=.07];
\draw [fill=white] (0.5,1) circle [radius=.07];
\draw [fill=black] (2,0.2) circle [radius=.07];
\draw [thick, red](-2,4)node[below left]{\large{$A$} } ;
\draw [thick, blue](2,4)node[below right]{\large{$B$} } ;
\end{tikzpicture}
\caption{\label{fig:PT1}}
\end{subfigure}\\
\begin{subfigure}[c]{.3\textwidth}
\begin{tikzpicture}[scale=0.5, decoration={markings, mark=at position 0.2 with {\arrow{>}}, mark=at position 0.4 with {\arrow{>}}, mark=at position 0.7 with {\arrow{>}}, mark=at position 0.9 with {\arrow{>}}}]

\draw[thick,->] (-2,0,0) -- (8,0,0) node[right]{$t$};
\draw[thick,->] (0,-2,0) -- (0,2,0) node[above]{$p$};
\draw[thick,->] (0,0,-2) -- (0,0,2) node[below left]{$q$};
\draw[postaction={decorate}, red, thick, domain=0:6, samples=10, smooth] plot (\x, {cos(deg(0.5*\x))^2},{-0.676263+3.38904*exp(-0.282177*\x)*sin(deg(0.823907*\x))}) node[above right]{$\Psi(t)$};
\foreach \i in {1.5,3,4.5} {
    \fill[red] (\i, {cos(deg(0.5*\i))^2},{-0.676263+3.38904*exp(-0.282177*\i)*sin(deg(0.823907*\i))}) circle (2pt);
}
\draw [fill=white] (0, {cos(deg(0.5*0))^2},{-0.676263+3.38904*exp(-0.282177*0)*sin(deg(0.823907*0))}) circle [radius=.07];
\fill[black] (6, {cos(deg(0.5*6))^2},{-0.676263+3.38904*exp(-0.282177*6)*sin(deg(0.823907*6))}) circle (2pt);
\draw[postaction={decorate}, blue, thick, domain=0:6, samples=10, smooth] plot (\x, {cos(deg(0.5*(6-\x)))^2},{-0.676263+3.38904*exp(-0.282177*(6-\x))*sin(deg(0.823907*(6-\x)))}) node[below right]{$\mathbb{T}\Psi(t)$};
\foreach \i in {1.5,3,4.5} {
    \fill[blue] (\i, {cos(deg(0.5*(6-\i)))^2},{-0.676263+3.38904*exp(-0.282177*(6-\i))*sin(deg(0.823907*(6-\i)))}) circle (2pt);
}
\draw [fill=white] (0, {cos(deg(0.5*(6-0)))^2},{-0.676263+3.38904*exp(-0.282177*(6-0))*sin(deg(0.823907*(6-0)))}) circle [radius=.07];
\fill[black] (6, {cos(deg(0.5*(6-6)))^2},{-0.676263+3.38904*exp(-0.282177*(6-6))*sin(deg(0.823907*(6-6)))}) circle (2pt);

\foreach \i in {0,1.5,3,4.5,6} {
    \fill[red, opacity=0.3] (\i,2,-2) -- (\i,2,-1) -- (\i,-2,-1) -- (\i,-2,-2) -- cycle;
    \fill[gray, opacity=0.1] (\i,2,-1) -- (\i,2,1) -- (\i,-2,1) -- (\i,-2,-1) -- cycle;
    \fill[blue, opacity=0.3] (\i,2,1) -- (\i,2,2) -- (\i,-2,2) -- (\i,-2,1) -- cycle;
}
\end{tikzpicture}
\caption{}
\end{subfigure}&
\begin{subfigure}[c]{.3\textwidth}
\begin{tikzpicture}[scale=0.5, decoration={markings, mark=at position 0.2 with {\arrow{>}}, mark=at position 0.4 with {\arrow{>}}, mark=at position 0.7 with {\arrow{>}}, mark=at position 0.9 with {\arrow{>}}}]

\draw[thick,->] (-2,0,0) -- (8,0,0) node[right]{$t$};
\draw[thick,->] (0,-2,0) -- (0,2,0) node[above]{$p$};
\draw[thick,->] (0,0,-2) -- (0,0,2) node[below left]{$q$};

\draw[postaction={decorate}, red, thick, domain=0:6, samples=10, smooth] plot (\x, {-cos(deg(0.5*\x))^2},{-0.177503 + 1.72611*exp(-0.128571*\x)*cos(deg(1.06524*\x))}) node[below right]{$\Psi(t)$};
\foreach \i in {1.5,3,4.5} {
    \fill[red] (\i, {-cos(deg(0.5*\i))^2},{-0.177503 + 1.72611*exp(-0.128571*\i)*cos(deg(1.06524*\i))}) circle (2pt);
}
\draw [fill=white] (0, {-cos(deg(0.5*0))^2},{-0.177503 + 1.72611*exp(-0.128571*0)*cos(deg(1.06524*0))}) circle [radius=.07];
\fill[black] (6, {-cos(deg(0.5*6))^2},{-0.177503 + 1.72611*exp(-0.128571*6)*cos(deg(1.06524*6))}) circle (2pt);
\draw[postaction={decorate}, blue, thick, domain=0:6, samples=10, smooth] plot (\x, {cos(deg(0.5*\x))^2},{0.177503-1.72611*exp(-0.128571*\x)*cos(deg(1.06524*\x))}) node[ right]{$\mathcal{P}\Psi(t)$};
\foreach \i in {1.5,3,4.5} {
    \fill[blue] (\i, {cos(deg(0.5*\i))^2},{0.177503-1.72611*exp(-0.128571*\i)*cos(deg(1.06524*\i))}) circle (2pt);
}
\draw [fill=white] (0, {cos(deg(0.5*0))^2},{0.177503-1.72611*exp(-0.128571*0)*cos(deg(1.06524*0))}) circle [radius=.07];
\fill[black] (6, {cos(deg(0.5*6))^2},{0.177503-1.72611*exp(-0.128571*6)*cos(deg(1.06524*6))}) circle (2pt);
\foreach \i in {0,1.5,3,4.5,6} {
    \fill[red, opacity=0.3] (\i,2,-2) -- (\i,2,-1) -- (\i,-2,-1) -- (\i,-2,-2) -- cycle;
    \fill[gray, opacity=0.1] (\i,2,-1) -- (\i,2,1) -- (\i,-2,1) -- (\i,-2,-1) -- cycle;
    \fill[blue, opacity=0.3] (\i,2,1) -- (\i,2,2) -- (\i,-2,2) -- (\i,-2,1) -- cycle;
}
\end{tikzpicture}
\caption{}
\end{subfigure}&
\begin{subfigure}[c]{.3\textwidth}
\begin{tikzpicture}[scale=0.5, decoration={markings, mark=at position 0.2 with {\arrow{>}}, mark=at position 0.4 with {\arrow{>}}, mark=at position 0.7 with {\arrow{>}}, mark=at position 0.9 with {\arrow{>}}}]

\draw[thick,->] (-2,0,0) -- (8,0,0) node[right]{$t$};
\draw[thick,->] (0,-2,0) -- (0,2,0) node[above]{$p$};
\draw[thick,->] (0,0,-2) -- (0,0,2) node[below left]{$q$};

\draw[postaction={decorate}, red, thick, domain=0:6, samples=10, smooth] plot (\x, {-cos(deg(0.5*(6-\x)))^2},{-0.177503 + 1.72611*exp(-0.128571*(6-\x))*cos(deg(1.06524*(6-\x)))}) node[below right]{$\Psi(t)$};
\foreach \i in {1.5,3,4.5} {
    \fill[red] (\i, {-cos(deg(0.5*(6-\i)))^2},{-0.177503 + 1.72611*exp(-0.128571*(6-\i))*cos(deg(1.06524*(6-\i)))}) circle (2pt);
}
\draw [fill=white] (0, {-cos(deg(0.5*6))^2},{-0.177503 + 1.72611*exp(-0.128571*6)*cos(deg(1.06524*6))}) circle [radius=.07];
\fill[black] (6, {-cos(deg(0.5*0))^2},{-0.177503 + 1.72611*exp(-0.128571*0)*cos(deg(1.06524*0))}) circle (2pt);
\draw[postaction={decorate}, blue, thick, domain=0:6, samples=10, smooth] plot (\x, {cos(deg(0.5*\x))^2},{0.177503-1.72611*exp(-0.128571*\x)*cos(deg(1.06524*\x))}) node[ right]{$\mathcal{PT}\Psi(t)$};
\foreach \i in {1.5,3,4.5} {
    \fill[blue] (\i, {cos(deg(0.5*\i))^2},{0.177503-1.72611*exp(-0.128571*\i)*cos(deg(1.06524*\i))}) circle (2pt);
}
\draw [fill=white] (0, {cos(deg(0.5*0))^2},{0.177503-1.72611*exp(-0.128571*0)*cos(deg(1.06524*0))}) circle [radius=.07];
\fill[black] (6, {cos(deg(0.5*6))^2},{0.177503-1.72611*exp(-0.128571*6)*cos(deg(1.06524*6))}) circle (2pt);
\foreach \i in {0,1.5,3,4.5,6} {
    \fill[red, opacity=0.3] (\i,2,-2) -- (\i,2,-1) -- (\i,-2,-1) -- (\i,-2,-2) -- cycle;
    \fill[gray, opacity=0.1] (\i,2,-1) -- (\i,2,1) -- (\i,-2,1) -- (\i,-2,-1) -- cycle;
    \fill[blue, opacity=0.3] (\i,2,1) -- (\i,2,2) -- (\i,-2,2) -- (\i,-2,1) -- cycle;
}
\end{tikzpicture}
\caption{}
\end{subfigure}
\end{tabular}
\caption{Examples of paths that initially fail to meet transition criteria but can be transformed into valid transition paths in the visiting ensemble  $\left\langle h_B\left(x_t\right)\right\rangle_{A B}^*$. (a) and (d) Show time reversal, (b) and (e) show parity reversal and (c) and (f) show combined parity-time reversal. \label{fig:Transformations}}
\end{figure*}
In this appendix, we will demonstrate how involutory (self-inverse) transformations can be used to enhance TPS methods under certain conditions. \Cref{fig:2-Way-Shooting-Sym} illustrates an abstract example of a trial trajectory, generated by two-way shooting, that initially fails to be reactive and a transformation $S$ that yields a reactive trajectory. \Cref{fig:Transformations} we shows some concrete examples of failed reactive paths and computationally cheap transformations that result in new reactive paths. (a) and (d) Show time reversal, (b) and (e) show parity reversal (space reflection) and (c) and (f) show combined parity-time reversal. In the Fock basis the action of the parity operator is simply given by 
\begin{equation}
    \mathbb{P} \ket{n}=(-1)^{n}\ket{n}. 
\end{equation}
In \cref{fig:MultiShotSym} we illustrate an example of mirror TPS. If we have one transformation operation, we generate two paths per shot. Considering the scenario in which a trajectory $\Psi^{(\mathrm{n})}$ is generated from by $\Psi^{(\mathrm{o})}$ by shooting and $S\Psi^{(\mathrm{n})}$ is generated from $\Psi^{(\mathrm{n})}$ by some transformation $S$. We then check if one or both of $\Psi^{(\mathrm{n})}$ and $S\Psi^{(\mathrm{n})}$ meet the endpoint constraints. If both meet the endpoint constraints we pick one at random with equal probability and record the number of solutions $n_s$. In the case both meet the endpoint criteria, we have
\begin{equation}
    n_s=2.
\end{equation}
Supposing $S\Psi^{(\mathrm{n})}$ is chosen, the metropolis acceptance rule is given by
\begin{multline}
\mathcal{P}_{\text {acc }}\left[\Psi^{(\mathrm{o})} \rightarrow\right. \left.S\Psi^{(\mathrm{n})}\right]= \\
\min \left\{1, \frac{\mathcal{P}\left[S\Psi^{(\mathrm{n})}\right] \mathcal{P}_{\mathrm{gen}}\left[S\Psi^{(\mathrm{n})} \rightarrow \Psi^{(\mathrm{o})}\right]}{\mathcal{P}\left[\Psi^{(\mathrm{o})}\right] \mathcal{P}_{\mathrm{gen}}\left[\Psi^{(\mathrm{o})} \rightarrow S\Psi^{(\mathrm{n})}\right]}\right\}.
\end{multline}
The stationary probabilities $\mathcal{P}(\Psi^{(\mathrm{o})})$ and $\mathcal{P}(S\Psi^{(\mathrm{n})})$ are given simply by
\begin{equation}
    \mathcal{P}(\Psi^{(\mathrm{o})})=\rho_{st}(\psi^{\mathrm{o}}_0)\prod_{i=0}^{N-1}\mathcal{P}(\psi^{\mathrm{o}}_{t_i}\to \psi^{\mathrm{o}}_{t_{i+1}}),
\end{equation}
and
\begin{equation}
    \mathcal{P}(S\Psi^{(\mathrm{n})})=\rho_{st}(S\psi^{\mathrm{n}}_0)\prod_{i=0}^{N-1}\mathcal{P}(S\psi^{\mathrm{n}}_{t_i}\to S\psi^{\mathrm{n}}_{t_{i+1}}).
\end{equation}
Since the choice both $\Psi^{(\mathrm{n})}$ and $S\Psi^{(\mathrm{n})}$ are generated from one shot we calculate the forward generation probability as
\begin{multline}
    \mathcal{P}_{\mathrm{gen}}\left[\Psi^{(\mathrm{o})} \to S\Psi^{(\mathrm{n})}\right]=\mathcal{P}_{\mathrm{gen}}\left[\Psi^{(\mathrm{o})} \to \Psi^{(\mathrm{n})}\right]\\
    =\frac{\mathcal{P}_{\mathrm{gen}}(\psi^{\mathrm{o}}_{t_s}\to\psi^{\mathrm{n}}_{t_s})}{n_s} \prod_{i=0}^{s-1} \bar{\mathcal{P}}(\psi^{\mathrm{n}}_{t_{i+1}}\to \psi^{\mathrm{n}}_{t_i})  \prod_{i=s}^{N-1}  \mathcal{P}(\psi^{\mathrm{n}}_{t_i}\to \psi^{\mathrm{n}}_{t_{i+1}}).
\end{multline}
For the reverse move generation probability we similarly have 
\begin{equation}
\mathcal{P}_{\mathrm{gen}}\left[S\Psi^{(\mathrm{n})}\to\Psi^{(\mathrm{o})}\right]=\mathcal{P}_{\mathrm{gen}} \left[S\Psi^{(\mathrm{n})}\to S\Psi^{(\mathrm{o})}\right],
\end{equation}
and thus
\begin{multline}\mathcal{P}_{\mathrm{gen}}\left[S\Psi^{(\mathrm{n})}\to\Psi^{(\mathrm{o})}\right] =\frac{\mathcal{P}_{\mathrm{gen}}(S\psi^{\mathrm{n}}_{t_{s'}}\to S\psi^{\mathrm{o}}_{t_{s'}})}{\bar{n}_s} \\ \prod_{i=0}^{s'-1} \bar{\mathcal{P}}(S\psi^{\mathrm{o}}_{t_{i+1}}\to S\psi^{\mathrm{o}}_{t_i})  \prod_{i=s'}^{N-1}  \mathcal{P}(S\psi^{\mathrm{o}}_{t_i}\to S\psi^{\mathrm{o}}_{t_{i+1}}),
\end{multline}
where $\bar{n}_s$ is the number of trajectories (either $S\Psi^{\mathrm{o}}$ or $\Psi^{\mathrm{o}}$) satisfying the endpoint constraints for the reverse move. If the shooting point displacement utilized to generate $\Psi^{(\mathrm{n})}$ from $\Psi^{(\mathrm{o})}$ is chosen from a symmetric distribution the ratio
\begin{equation}
    \frac{\mathcal{P}_{\mathrm{gen}}(S\psi^{\mathrm{n}}_{t_{s'}}\to S\psi^{\mathrm{o}}_{t_{s'}})}{\mathcal{P}_{\mathrm{gen}}(\psi^{\mathrm{o}}_{t_s}\to\psi^{\mathrm{n}}_{t_s})}=\frac{\mathcal{P}_{\mathrm{gen}}(-S\delta p)}{\mathcal{P}_{\mathrm{gen}}(\delta p)}=1
\end{equation}
is unity due to $S$ being an involutory transformation which can only change the sign of $\delta P$ and not the magnitude. For the acceptance probability we have 
\begin{multline} \label{eq:probMultiShot}
\mathcal{P}_{\text {acc }}\left[\Psi^{(\mathrm{o})} \rightarrow\right. \left.S\Psi^{(\mathrm{n})}\right]= \\\frac{n_s}{\bar{n}_s}\frac{\rho_{st}(S\psi^{\mathrm{n}}_0)\prod_{i=0}^{N-1}\mathcal{P}(S\psi^{\mathrm{n}}_{t_i}\to S\psi^{\mathrm{n}}_{t_{i+1}})}{\rho_{st}(\psi^{\mathrm{o}}_0)\prod_{i=0}^{N-1}\mathcal{P}(\psi^{\mathrm{o}}_{t_i}\to \psi^{\mathrm{o}}_{t_{i+1}})}\\
\frac{\prod_{i=0}^{s'-1} \bar{\mathcal{P}}(S\psi^{\mathrm{o}}_{t_{i+1}}\to S\psi^{\mathrm{o}}_{t_i})  \prod_{i=s'}^{N-1}  \mathcal{P}(S\psi^{\mathrm{o}}_{t_i}\to S\psi^{\mathrm{o}}_{t_{i+1}})}{\prod_{i=0}^{s-1} \bar{\mathcal{P}}(\psi^{\mathrm{n}}_{t_{i+1}}\to \psi^{\mathrm{n}}_{t_i})  \prod_{i=s}^{N-1}  \mathcal{P}(\psi^{\mathrm{n}}_{t_i}\to \psi^{\mathrm{n}}_{t_{i+1}})}.
\end{multline}
For $S=\mathbb{T}$ corresponding to time reversal, $t_{s'}=t_{N-s}$ the above acceptance probability reduces to
\begin{multline} \label{eq:probTShot}
\mathcal{P}_{\text {acc }}\left[\Psi^{(\mathrm{o})} \rightarrow\right. \left.\mathbb{T}\Psi^{(\mathrm{n})}\right]= \\\frac{n_s}{\bar{n}_s}\frac{\rho_{st}(\mathbb{T}\psi^{\mathrm{n}}_0)\prod_{i=s}^{N-1}\bar{\mathcal{P}}(\psi^{\mathrm{n}}_{t_{i+1}}\to \psi^{\mathrm{n}}_{t_{i}})}{\rho_{st}(\psi^{\mathrm{o}}_0)\prod_{i=s'}^{N-1}\mathcal{P}(\psi^{\mathrm{o}}_{t_i}\to \psi^{\mathrm{o}}_{t_{i+1}})}
\frac{ \prod_{i=s'}^{N-1}  \mathcal{P}(\psi^{\mathrm{o}}_{t_{i+1}}\to \psi^{\mathrm{o}}_{t_{i}})}{  \prod_{i=s}^{N-1}  \mathcal{P}(\psi^{\mathrm{n}}_{t_i}\to \psi^{\mathrm{n}}_{t_{i+1}})}.
\end{multline}
For arbitrary $S$ however, the acceptance rule \eqref{eq:probMultiShot} may lack efficient cancellation of forward or backward segments, leading to a longer evaluation time. Consequently, there is a need to strike a balance between the decreased time spent on integrating dynamics due to the increased probability of meeting endpoint constraints and the increased time required for calculating acceptance probabilities. The mirror method can present notable advantages in scenarios with a high number of spatial dimensions. In such cases, it offers the benefit of an additional parity transformation for each dimension, potentially enhancing efficiency.

\end{document}